\documentclass[prfluids,onecolumn,showpacs,oneside]{revtex4-2}

\usepackage{graphicx}  
\usepackage{subfigure}
\usepackage{multirow}
\usepackage{fancyhdr}
\usepackage{longtable}
\usepackage{parskip}
\usepackage{tikz}
\usepackage{cancel}
\usepackage{silence}
\usepackage{hyperref}
\usepackage[T1]{fontenc}
\usepackage{dcolumn}   

\usepackage{bm}        
\usepackage{amsfonts}  
\usepackage{amsmath}   
\usepackage{amssymb}   
\usepackage{cleveref} 
\usepackage{layouts}   
\usepackage[paperwidth=210mm,paperheight=297mm,centering,hmargin=0.8 in,vmargin=1 in]{geometry}
\usepackage[bottom]{footmisc}

\usepackage[pscoord]{eso-pic}

\newcommand{\hst}{{h^*}}
\newcommand{\rst}{{r^*}}
\newcommand{\ost}{{\Omega^*}}

\begin{document}
\title{A Swimming Rheometer: Self-propulsion of a freely-suspended swimmer enabled by viscoelastic normal stresses}

\date{\today}

\author{Laurel A. Kroo$^{* 1}$, Jeremy P. Binagia$^{* 2}$, Noah Eckman$^2$, Manu Prakash$^3$, Eric S. G. Shaqfeh$^{1,2}$ \\(\small *Equal Contribution)}

\affiliation { \vspace{7pt} $^1$Department of Mechanical Engineering, Stanford University}
\affiliation { $^2$Department of Chemical Engineering, Stanford University}
\affiliation { $^3$Department of Bioengineering, Stanford University}

\begin{abstract}  

Self-propulsion at low Reynolds number is notoriously restricted, a concept that is commonly known as the "scallop theorem". Here we present a truly self-propelled swimmer (force- and torque-free) that, while unable to swim in a Newtonian fluid due to the scallop theorem, propels itself in a non-Newtonian fluid as a result of fluid elasticity. This propulsion mechanism is demonstrated using a robotic swimmer, comprised of a "head" sphere and a "tail" sphere, whose swimming speed is shown to have reasonable agreement with a microhydrodynamic asymptotic theory and numerical simulations. Schlieren imaging demonstrates that propulsion of the swimmer is driven by a strong viscoelastic jet at the tail, which develops due to the fore-aft asymmetry of the swimmer. Optimized cylindrical and conic tail geometries are shown to double the propulsive signal, relative to the optimal spherical tail. Finally, we show that we can use observations of this robot to infer rheological properties of the surrounding fluid. We measure the primary normal stress coefficient, $\Psi_1$ at shear rates $< 1$ Hz, and show reasonable agreement with extrapolated bench-top measurements (between $0.8-1.2$ Pa sec$^2$ difference).We also discuss how our swimmer can be used to measure the second normal stress coefficient, $\Psi_2$, and other rheological properties. The study experimentally demonstrates the exciting potential for a “swimming rheometer”, bringing passive physics-driven fluid sensing to numerous applications in chemical and bioengineering.   
\end{abstract}


\maketitle 

 \section*{Introduction} 
We consider the design of a swimming robot that passively updates its observable behavior to reflect attributes of its environment.
Indeed, bio-mimetic adaptation and sensing are advantageous in improving robustness and redundancy for a wide-range of robotic systems in unknown environments \cite{webb2001slocum} \cite{parness2009microfabricated}. Within nature itself, microorganisms must adapt to and sense complex fluid environments to, among other things, escape from predators, forage for food, and to infect their host \cite{spagnolie2015complex} \cite{gilpin2017vortex} \cite{bull2021excitable}. For example, the bacteria \textit{H. pylori}, known for causing stomach ulcers, must first traverse a thick layer of gastric mucus before it can cause infection \cite{celli2009helicobacter}. Owing to their microscopic size and the very viscous fluids in which they are commonly immersed, these microorganisms live in a world where fluid inertia is effectively non-existent, i.e. they are said to swim at truly zero Reynolds number \citep{purcell1977life}. As Purcell described in his famous talk "Life at low Reynolds number", the fact that microswimmers move at $\mathrm{Re} = 0$ sets extraordinary constraints on the types of swimming strategies, or "gaits", they can adopt. These constraints all originate from the fact that the Stokes equations (i.e. the governing equations for Stokes flow) are linear, in contrast to the highly nonlinear Navier-Stokes equations that govern flows with inertia \citep{kim2013microhydrodynamics}. For example, it is well known \citep{purcell1977life} that swimmers trying to move via reciprocal motion in Stokes flow, wherein the swimming stroke is identical when viewed forwards and backwards in time, exhibit zero net propulsion since any progress they make in the first half of their stroke will be exactly cancelled by translation equal in magnitude but opposite in direction in the latter half. Purcell's elegant example of this is a scallop that tries to swim by simply opening and closing its shell. Because the scallop's gait exhibits time-reversal symmetry, it exhibits zero net motion, a result now widely known as the Scallop Theorem \citep{purcell1977life}. 

\begin{figure}[ht!]
    \centering
    \includegraphics[width = \textwidth]{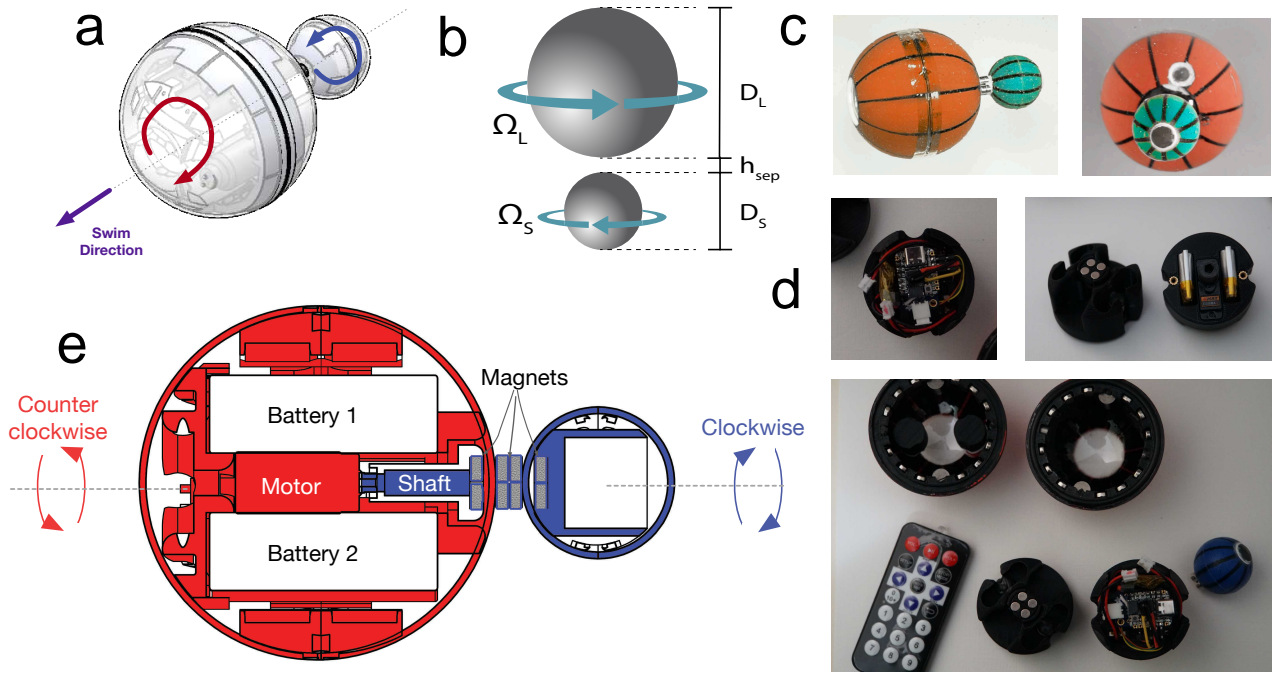}

    \caption{(a) The design of a rotationally-symmetric swimming robot is pictured. Note that both the head and tail are spheres of different sizes. (b) Illustration of the model two-sphere swimmer considered in the asymptotic theory and numerical simulations. The swimmer consists of a large and small sphere, each having diameter $D_L$ and $D_S$ respectively. The dimensional distance between the two spheres is denoted by $h_{sep}$ and the magnitude of the rotation rates for each sphere are $\Omega_L$ and $\Omega_S$. In \cref{sec:math}, the axis of revolution or the vertical direction in this figure is taken to be the $z$ direction (imagining a cylindrical coordinate system centered at the center of the swimmer). (c) We have built a small robot with a head diameter of 70mm and a tail diameter of 30mm, pictured here from the side and rear, while submerged in an elastic fluid. (d) The robot, pictured here deconstructed, contains a small microcontroller, photodiode, stepper motor, motor controller, batteries, voltage regulator and gearbox. The device is optically-controlled via a modified IR remote (850nm). A gasket at the equator is compressed by a series of magnets on each hemisphere. Ballast tanks allow for fine mass tuning. (e) a motor placed in the head drives the system to rotate. The motor rotates a drive shaft within the head, which is magnetically coupled to the tail. This allows the distance between the head and tail to be reconfigurable, and the waterproofing to be robust.The device is controlled via an adapted IR remote control. It is battery powered and has closed-loop speed control. A gearbox is used to achieve very slow, continuous rotation speeds.}
    \label{fig:fig-1-schematic}
\end{figure}

To overcome the aforementioned constraints, real microorganisms utilize geometric asymmetries or introduce other nonlinearities into the problem (e.g. having a flexible rather than rigid tail) to enable propulsion at $\mathrm{Re} = 0$ \cite{lauga2009hydrodynamics}. One interesting example of this is propulsion that is enabled by nonlinearities present in the surrounding fluid itself. For example, if the surrounding fluid is viscoelastic, meaning it responds both viscously like a liquid and elastically like a solid when deformed, then propulsion mechanisms forbidden in Newtonian Stokes flow suddenly become possible in this non-Newtonian fluid. For example, \citet{qiu2014swimming} demonstrated experimentally that the scallop shape discussed by Purcell actually can swim when placed in non-Newtonian (shear-thinning and shear-thickening) fluids. Similar propulsion enabled by non-Newtonian rheology (in particular, fluid elasticity) was also demonstrated independently by \citet{keim2012fluid} and \citet{datt2018two} through reciprocal motion of an artificial "dimer" and "two-sphere" swimmer respectively. Another example is the work by \citet{pak2012micropropulsion} and \citet{puente2019viscoelastic}, where any object consisting of two spheres of unequal sizes (resembling a snowman) is made to rotate in a viscoelastic fluid through the use of an external magnetic field. They find through theory, simulations, and experiments that such an object exhibits zero propulsion in a Newtonian fluid at $\mathrm{Re} = 0$ but is able to break symmetry and propel in the direction of the smaller sphere when placed in a viscoelastic fluid. However, in this case the object is not a freely-suspended swimmer.

We present a robotic swimmer that not only leverages non-Newtonian rheology to propel itself in a viscoelastic fluid, but it does so as a freely suspended swimmer. This swimmer consists of two counter-rotating spheres of unequal sizes; the relative rotation rate of the two spheres is prescribed via an on-board microcontroller as opposed to the body being rotated rigidly via an external field. Because our swimmer is thus both force- and torque-free, we believe it can more faithfully capture the swimming dynamics of real microswimmers. An additional benefit to truly torque-free designs is that these devices are highly portable, requiring little to no infrastructure to operate. Because there is no external field that could add error to the system (or that the device could swim away from), temporal changes in the response of the swimmer are reflective of changes in the environment, such as spatial gradients in fluid properties.   
Finally, as will be shown, the rotation of the head and the forward propulsion speed of the swimmer are directly related to important rheological properties of the surrounding fluid. The device is therefore not only an interesting model of a low Reynolds number swimmer, but it also could be used as a portable, rheological probe in different fluid environments.  

\section*{Robot Design and Experiments}
A small robot was constructed that is capable of untethered self-propulsion in viscoelastic fluids (fig 1b, 1c). The robot was designed in two modules: the head (70mm diameter) and tail (30mm diameter). The head contains all power and electronics for the system, including a pseudo-closed-loop speed control system consisting of a small stepper motor and gearbox. The initial versions of the robot were constructed within molded polypropylene spherical shells to ensure high geometric precision of the exterior, to match with simulations and theory. Both the head and tail modules required ballast tanks for tuning the buoyancy and center of mass of the robot. These tanks were accessible from the exterior, and use rubber "vacutainer" seals - which allow the user to repeatably inject or withdraw fluid from the ballasts at multiple points on the robot (see: \cref{sec:schlieren}).  

The head and tail are coupled with a magnetic linker shaft (thus, the design includes no rotary shaft seal). This allows the robot to be highly configurable and modular, specifically allowing for quickly swapping tail attachments without any modification to the robot. Because the system has closed-loop speed control (and is not power-limited), effects of friction in the magnetic linker do not significantly affect the system performance.

The device is powered with small lithium polymer batteries, with an onboard voltage regulator, microcontroller, IR receiver and stepper motor driver. Speed control protocols can be uploaded to the onboard microcontroller, and transmitted during experiments with a modified IR remote control (transmitter modified from 940nm to 850nm to avoid significant absorption from the submersion fluid). The robot is capable of receiving signal at significant distances while fully submerged (tested to approximately 0.5 meters without receiver difficulty). 

The swimmer was tested in several fluids. Particularly, we confirmed that in Newtonian control fluids such as water and corn syrup, the swimmer does not propel itself. To test the swimmer in viscoelastic fluid, approximately 3 gallons of a polyacrylamide-based Boger fluid was mixed (see: \cref{sec:rheology}). This viscoelastic fluid was designed specifically to have a large first normal stress coefficient, $\Psi_1$, and minimal shear-thinning for the initial proof of concept experiments. Rheological properties of the fluid were characterized on a benchtop rheometer for use in the numerical simulations and asymptotic theory (see: \cref{sec:rheology}). 

    
    
    


\section*{Summary of asymptotic theory and numerical simulations}
\label{sec:summary-math}

While we defer the detailed derivation of the asymptotic theory and numerical simulations to the Methods section (\cref{sec:math}) and prior work (i.e. \citep{binagia2021self}), we briefly summarize them below. To model the motion of our robotic swimmer via theory and numerical simulations, we adopt the model illustrated in \cref{fig:fig-1-schematic}b. The model swimmer consists of two spheres of unequal size, whose diameters are $D_L$ and $D_S$ respectively. The spheres are separated by a distance $h_{sep}$ and the magnitude of the rotation rates for the two spheres are $\Omega_L$ and $\Omega_S$ respectively. Note that the angular velocities of the two spheres must necessarily be in opposite directions to maintain the torque-free condition on the swimmer, as discussed in further detail below. The distance between the centers of the two spheres is given by $h^{'}=R_L+h_{sep}+R_S$ where $R_L$ and $R_S$ are the radii of the two spheres. 

The motion of the fluid is governed by conservation of momentum and mass at zero Reynolds number, written in dimensionless form as: 
\begin{equation}
    \nabla\cdot
    \bm{\sigma}=0, \quad\quad \nabla\cdot\mathbf{u}=\mathbf{0}
\end{equation}
where $\mathbf{u}$ is the velocity of the fluid, and $\bm{\sigma}$ is the Cauchy stress. Note that both the equations above and those that follow are made dimensionless by scaling lengths by the radius of the large sphere $R_L$, time by the inverse of the smaller sphere's rotation rate, $\Omega_S^{-1}$, velocities by the product $R_L \Omega_S$, and stresses by $\mu_0 R_L^2 \Omega_S$. Because the fluid is viscoelastic, we write the total stress $\bm{\sigma}$ as the sum of a Newtonian and polymeric contribution: 
\begin{equation}
    \bm{\sigma} = -p\mathrm{\mathbf{I}} + \beta(\nabla\mathbf{u} + \nabla\mathbf{u}^T) + \bm{\tau}^p
\end{equation}
where $p$ denotes pressure, $\bm{\tau}^p$ is the extra stress coming from deformation of polymer molecules immersed in the fluid, and $\beta=\mu_s/(\mu_s+\mu_p)=\mu_s/\mu_0$ is the ratio of the solvent viscosity $\mu_s$ to the total zero-shear solution viscosity $\mu_0$ (for a fluid with polymer viscosity equal to $\mu_p$). In general, we use the Giesekus constitutive equation to describe the extra polymer stress $\bm{\tau}^p$:
\begin{align}
&\bm{\tau}^{p} = \frac{1-\beta}{\text{De}}(\mathbf{c}-\mathbf{I}) \label{eq:mt-giesekus-1} \\
&\text{De} \stackrel{\triangledown}{\mathbf{c}} + (\mathbf{c}-\mathbf{I}) + \alpha_m(\mathbf{c}-\mathbf{I})^2   = \mathbf{0}.
\label{eq:mt-giesekus}
\end{align}
In the above equation, $\mathbf{c}$ is the conformation tensor and  $\stackrel{\triangledown}{\mathbf{c}}=\partial \mathbf{c}/\partial t+\mathbf{u}\cdot\nabla\mathbf{c}-\nabla\mathbf{u}^T\cdot\mathbf{c}-\mathbf{c}\cdot\nabla\mathbf{u}$ is the upper-convected derivative. $\mathrm{De}=\lambda \Omega_S$ is the Deborah number, which describes the relative importance of elastic effects in a viscoelastic fluid with relaxation time $\lambda$. 
If the Giesekus mobility parameter $\alpha_m$ is chosen to be equal to zero, the simpler Oldroyd-B model is recovered, which models the polymer molecules in the fluid to be Hookean dumbbells \citep{oldroyd1950formulation}. Nonzero values of $\alpha_m$ allow the Giesekus equation to model drag anisotropy experienced by the Hookean dumbbells; this permits the model to predict more realistic rheological behavior such as shear-thinning and nonzero second normal stress differences \citep{bird1987dynamics}. As discussed in the Methods section, this formulation can be readily extended to a multi-mode description of the polymer stress, whereby each mode has its own relaxation time $\lambda^{(i)}$ and viscosity $\mu^{(i)}_p$ and whose individual stress is given by \cref{eq:mt-giesekus-1,eq:mt-giesekus}; the total extra polymer stress in this case is then just the sum of the stress contributed by each mode. 
This set of governing equations is solved alongside an appropriate set of boundary conditions (no-slip at the surface of the swimmer) and constraints (i.e. that the swimmer is net force and torque free \citep{lauga2009hydrodynamics}) both numerically and through a far-field asymptotic theory valid for small De. The details of the numerical solution methodology are listed in \cref{sec:simulations} and prior work (\citep{binagia2021self}). In brief, the numerical solution considers the co-moving frame of reference such that a body-fitted mesh may be considered around the swimmer (with increasing mesh resolution in the vicinity of the swimmer). The flow solver used is a higly-accurate (third-order) finite volume flow solver developed at Stanford \citep{Ham2006} that has been validated for a wide range of problems, including viscoelastic flows \citep{Richter2010,padhy2013simulations,yang2016numerical}, deformable particles \citep{saadat2018immersed}, and active swimmers \citep{binagia2019three,binagia2020swimming,housiadas_binagia_shaqfeh_2021}. The swimmer's speed $U$ and head rotation rate $\Omega_L$ are determined through an iterative process at each time step (using Broyden's method \citep{broyden1965class}) such that the swimmer is force- and torque-free.
Alongside our numerical solution is a far-field asymptotic analysis valid for small De. The details may be found in \cref{sec:theory-section} and \citet{binagia2021self}, while the key results are summarized below. It is shown using a far-field asymptotic analysis in \cref{sec:theory-section} that for small De the speed of the swimmer is of the general form:
\begin{equation}
    U = \mathrm{De_{SO}}(1-B)f(\rst,\hst,\ost)
    \label{eq:u-summary-eq}
\end{equation}
where $\mathrm{De_{SO}}=\mathrm{De}(1-\beta)$, $B=-2\Psi_2/\Psi_1$. Here $\Psi_1$ and $\Psi_2$ are the first and second normal stress coefficients of the fluid, i.e. rheological parameters that characterize the elasticity of the fluid. $f(\rst,\hst,\ost)$ is a function that depends on the particular geometry and kinematics of the swimmer: $\rst=R_S/R_L$ is the ratio of the radii of the two spheres, $\hst=h_{sep}/R_L$ is the dimensionless distance between the two spheres, and $\ost=\Omega_S/\Omega_L$ is the ratio of their respective rotation rates. We refer the reader to \cref{eqn:uz-split,eqn:uz0,eqn:uzHO} in the Methods section for the full expression of $U$ written explicitly in terms of $\rst$, $\hst$, and $\ost$.  
Collectively, the viscosity $\mu_0$ and the normal stress coefficients $\Psi_1$ and $\Psi_2$ are referred to as the viscometric functions of a fluid \citep{bird1987dynamics}. They are defined with respect to a steady shear flow, i.e. 
\begin{align}
    &\tau_{yx} = \mu_0 \dot\gamma \\ 
    &\tau_{xx}-\tau_{yy}=\Psi_1 \dot\gamma^2 \\
    &\tau_{yy}-\tau_{zz}=\Psi_2 \dot\gamma^2
\end{align}
where in the above the direction of the shear flow is taken to be in the positive x-direction and $\dot\gamma$ is the shear rate. For a Newtonian fluid, where there are no differences in the normal stresses, $\Psi_1=\Psi_2=0$. In contrast, for a viscoelastic fluid, $\Psi_1$ and $\Psi_2$ are nonzero, a fact that leads to unusual flow behavior; the most notable of these is the Weissenberg rod-climbing effect, whereby fluid climbs up a rod rotating in an elastic fluid as a result of elastic hoop stresses \citep{boger2012rheological}. Note that for most polymeric fluids $\Psi_1$ is positive and $\Psi_2$ is small and negative \citep{bird1987dynamics}.
We can rewrite \cref{eq:u-summary-eq} such that the first and second normal stress coefficients of the fluid, $\Psi_1$ and $\Psi_2$, appear as dependent variables:
\begin{equation}
    \Psi_1\bigg(1+\frac{2\Psi_2}{\Psi_1}\bigg) = \frac{2 U' \mu_0}{\Omega_S^2 R_L f(\rst,\hst,\ost)}
    \label{eq:first-eq}
\end{equation}
where $U'$ is the dimensional swimming speed. From this equation, one can clearly see how the elastic properties of the fluid ($\Psi_1$ and $\Psi_2$) can be determined from a measurement of the robot's swimming speed $U'$ for a prescribed tail rotation rate $\Omega_S$. We note that for most polymeric fluids of interest $|\Psi_2|/\Psi_1 \ll 1$ (a common heuristic is $\Psi_2=-0.1 \Psi_1$) \citep{bird1987dynamics,morrison2001understanding}, and so the left-hand side of \cref{eq:first-eq} simplifies to just $\Psi_1$. Thus, assuming $\Psi_2 \approx 0$, one can directly measure $\Psi_1$ for a fluid so long as the viscosity $\mu_0$ is known. Alternatively, and as we will be described in the discussion section, one can also determine both $\Psi_1$ and $\Psi_2$ simultaneously by measuring both the swimmer's speed $U'$ and the rotation rate of the head (which itself will depend on the rheology of the fluid) for a single experiment. 

\section*{Results}  
\subsection*{Characterization of Propulsion Speed}

\begin{figure}[ht!]
    \centering
    \includegraphics[width=1.0\textwidth]{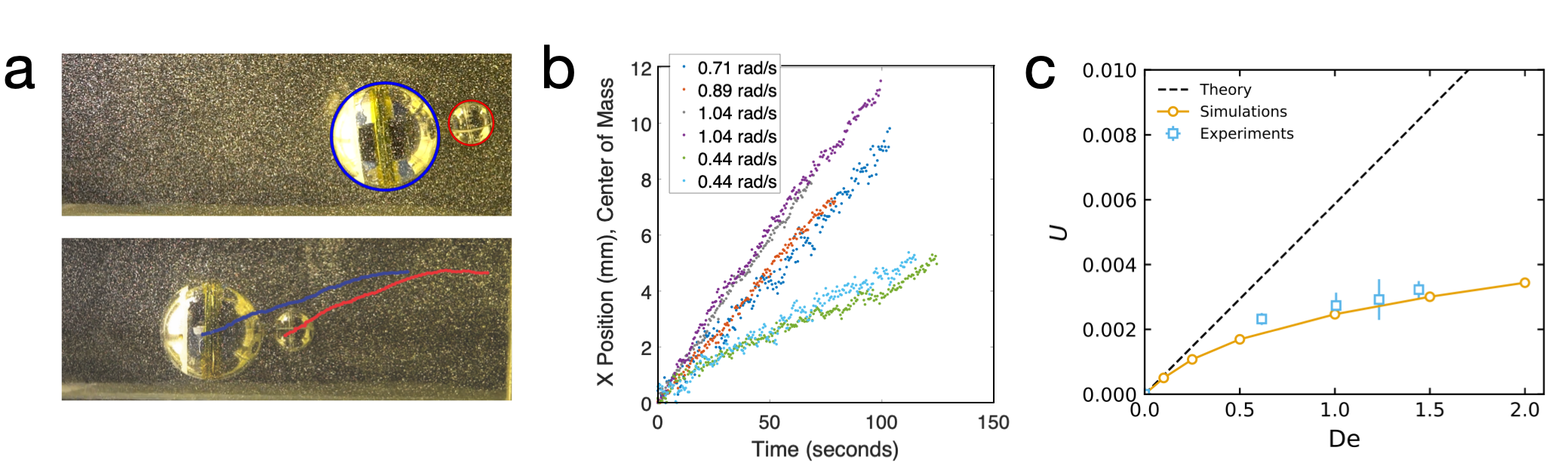}
    \caption{(a) Swimmers were submerged in elastic fluid, and image analysis was performed to determine propulsion speed. Pictured above is the head and tail identification, and below, the trajectories in space of the head (blue) and tail (red), respectively. (b) For different tail rotation speeds, the coordinates of the center of mass were tracked. Here we show x position vs. time data during steady-state swimming at different rotation rates. (c) Dimensionless swimming speed $U=U'/(\Omega_S R_L)$ as a function of the Deborah number, De. The analytical theory, valid for small De, is shown as the dotted black line and assumes an infinite domain and zero shear-thinning in the fluid. This curve was generated using \cref{eqn:uz-split,eqn:uz0,eqn:uzHO} with the rheological parameters determined from a fit to the multi-mode Oldroyd-B equation (the values of which are listed in \cref{tab:my-table}). Experimental data is shown as the blue square markers, with error bars denoting the standard deviation of measurements over repeated trials. Simulations (orange circles) are performed with a multi-mode Giesekus model (which allows for shear-thinning and second normal stress differences in the fluid), with the mesh domain defined to approximately match the size of the experimental tank (the confinement ratio $C=2R_L/R_{domain}=0.32$ where $R_{domain}$ is the radius of our cylindrical computational domain). The theory and simulations take as input the geometry of the robotic swimmer and the rheology of the fluid: $\hst=0.17$, $\rst=0.43$, $\bar{\lambda}=1.38$ s and $\beta=0.46$.}
    \label{fig:u-vs-de}
\end{figure}

As described in the preceding section, the swimming speed of the robot is directly related to the elasticity of the fluid. It is critical then to assess how the speed of the robotic swimmer compares to that predicted by theory and simulations since the latter two may then be used to infer properties of the fluid based on observed swimming speeds. In \cref{fig:u-vs-de}c, the dimensionless swimming speed of the robot $U=U'/(\Omega_S R_L)$ is plotted as a function of the Deborah number (De) of the fluid. De$ =\bar{\lambda} \Omega_S$ where $\bar{\lambda}=(\sum_{i=0}^n \lambda^{(i)} \mu_p^{(i)})/(\sum_{i=0}^n \mu_p^{(i)})$ is the average relaxation time for a fluid described with a spectrum of relaxation times (i.e. using a multi-mode rheological model). Experimental data, shown as the blue squares in \cref{fig:u-vs-de}c, is plotted alongside our small De asymptotic theory (black dashed line) and our numerical simulations (orange markers). To compare to experiments, both the theory and simulations take as input the specific geometry of the robotic swimmer and the rheological characterization of our PAA-based Boger fluid (c.f. \cref{fig:u-vs-de} for more details).  

From \cref{fig:u-vs-de}c, we observe that both experiments and simulations predict a modest increase in dimensionless swimming speed as a function of the Deborah number De. The asymptotic theory agrees with numerical simulations for small De, but the two depart as De increases, likely as a result of the fact that the theory does not consider any amount of shear-thinning in the fluid nor confinement effects, both of which have been shown to diminish the swimming speed for larger values of De (see \cref{sec:conf-and-st}). Note that the disagreement of the theory for large De is to be expected, since this theory was derived in the limit of De $\ll 1$. The relative agreement between simulations and experiments serve as a validation for our computational model and thus allow us to make claims about the propulsion mechanism through an analysis of the numerical results.


\subsection*{Propulsion Mechanism}

The propulsion mechanism of this device is directly related to the first normal stress coefficient, as predicted in previous theoretical work \citep{binagia2021self}. This is further supported by the experimental observation that the swimmer cannot propel itself in Newtonian, non-elastic fluids at low Reynolds numbers. However, it is not immediately clear how normal stresses in the fluid are mechanistically generating swimmer thrust.

\begin{figure}[ht!]
    \centering
    \includegraphics[width=\linewidth]{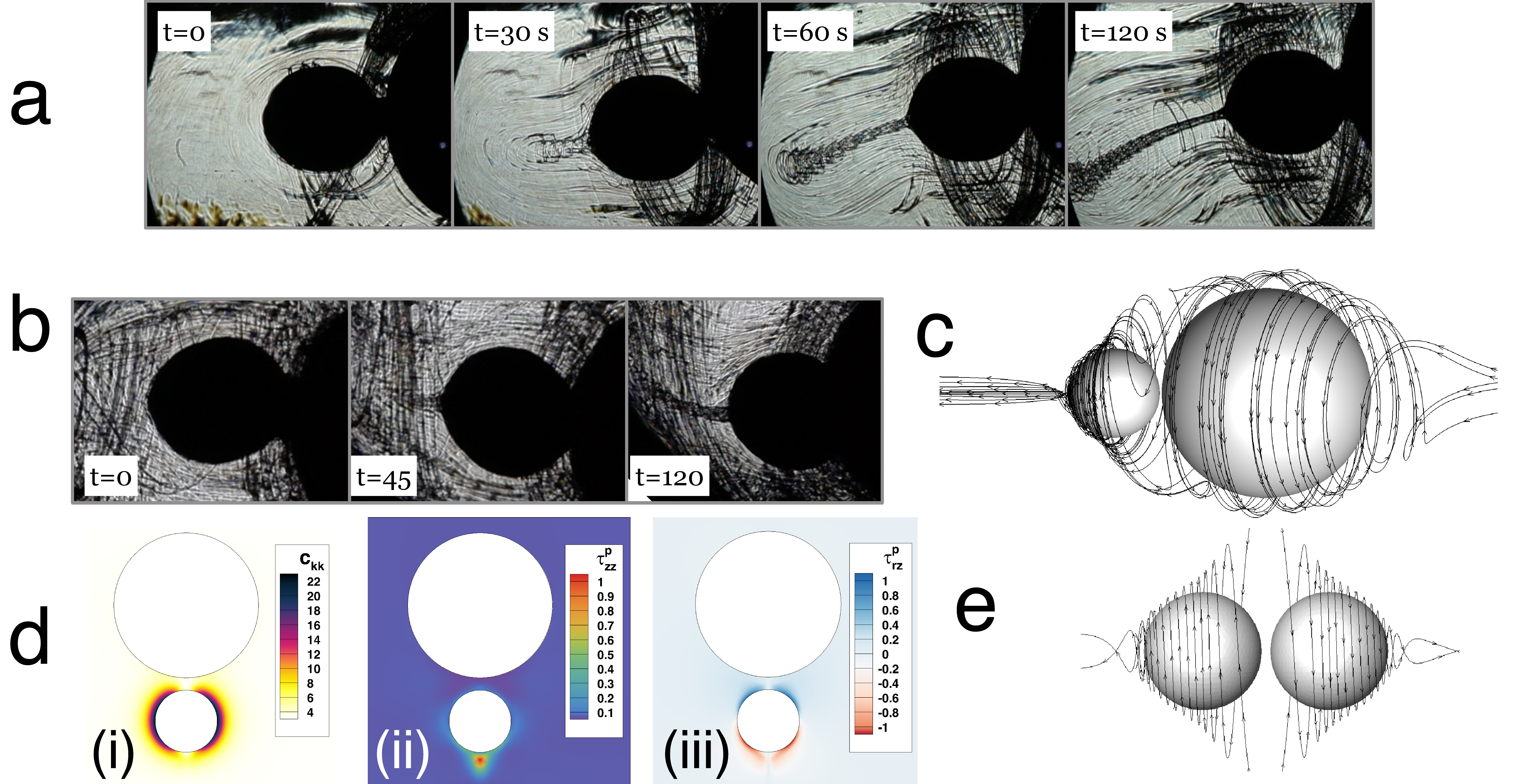}
    \caption{(a) Schlieren imaging was used to image features of the flow field in experiments. Shortly after the rotation of the small sphere is initiated ($t=0$, $t=30$), we observe the formation of hoop-like strands around the small sphere and the creation of a slender jet-like structure at the pole of the small spherical tail. As time progresses ($t=60$), the jet is seen to persist and grow in length. Even at long times ($t=120$), this jet structure is still present behind the rear of the swimmer. (b) We have repeated this imaging process with multiple trials and observe the formation of this jet each time. (c) Through visualizing the streamlines of the flow via numerical analysis, we can observe that a similar elastic jet is present in the numerical simulations using the single-mode Giesekus equation ($\beta=0.46$, $\mathrm{De}=2$, $\alpha_m=0.035$, $\rst=0.43$, $\hst=0.17$) (d) Numerical simulations indicate that hoop stresses are dominant (i) but that net polymer stresses do exist in along the direction of the robot's propulsion (ii) and in the $\theta$-z stress component behind the tail (iii). (e) We have shown with numerical simulations that when the head and tail are identical in size, the jet structures appear both at the head and tail.}
    \label{fig:jetformation}
\end{figure}

To further investigate, we imaged the robot using a slightly modified Schlieren imaging set-up (discussed further in \cref{sec:schlieren}). As shown in \cref{fig:jetformation}a and \cref{fig:jetformation}b, features of the flow indicate that a narrow viscoelastic jet develops behind the tail of the robot over a period of tens of seconds. Significant flow in the azimuthal direction (i.e. around the spinning tail) is also visable as dark strands around the small sphere; such flow is predicted to generate regions of large hoop stress around the tail of the swimmer \citep{binagia2021self}.

Numerical simulations indicate a set of fluid structures that match those seen in the experiments (see: \cref{fig:jetformation}c). Simulations suggest that while hoop stresses dominate the polymer stress, the front-back asymmetry of these stresses along the swimming axis leads to a region of high pressure behind the tail (see: \cref{fig:jetformation}d), and a non-negligible polymer stress develops along the swimming direction ($\tau_{zz}^p$, see: \cref{fig:jetformation}d) in the region of the viscoelastic jet.

While the elastic properties of the fluid enable the robot to propel itself, it is the interaction between the solid bodies (specifically, in the context of an  asymmetry between the head and the tail) and the nonlinearity of this viscoelastic fluid that enable it to form a primary jet and region of high pressure behind the tail. \cref{fig:jetformation}e demonstrates this concept, showing that when the head and tail are of equal size, there exist two such jets also of equal size, and each expel fluid from the head and tail, with no net propulsion. Fundamentally, the formation of each jet is a manifestation of the Weissenberg effect (or the so-called "rod-climbing effect") that was discussed previously in the introduction \citep{boger2012rheological,bird1987dynamics}.

The propulsion mechanism is analogous to the manner in which a wing produces lift, in that there are two independent critical factors: body shape and a nonlinearity in the equations of motion. Lift is a body force that requires the generation of a fluid structure (circulation). Without the non-linearity of inertia in the momentum balance equations, circulation is exceptionally difficult to generate. But even for highly inertial flows, the shape of the airfoil is critical in generating lift. 
Similarly in the viscoelastic propulsion mechanism described above, the combination of a nonlinear constitutive equation (including fluid elasticity) and a specific body shape (requiring head / tail asymmetry), leads to the generation of a fluid structure (i.e. a single dominant viscoelastic jet). This fluid structure gives rise to the net propulsive effect we observe over a variety of operating conditions in \cref{fig:u-vs-de}. 
In the context of viscoelastic fluids, this naturally leads to the question of optimizing this propulsive effect, by the tuning the swimmer's geometry under different operating conditions. 
\subsection*{Geometric Optimization}
In the interest of maximizing the propulsion of our robotic swimmer, we conducted numerical simulations for a range of tail shapes and geometries to ascertain the shapes that maximize the swimming speed. We focused specifically on the dimensionless swimming speed $U$ since the speed is the primary signal used to infer the rheological parameters of the fluid when using the robot as a fluid sensor. In other words, by maximizing the speed $U$ for a given tail rotation rate $\Omega_S$ and fluid rheology, one maximizes the signal-to-noise ratio for this robot as a measurement device.

\begin{figure}[ht!]
    \centering
    \includegraphics[width=\linewidth]{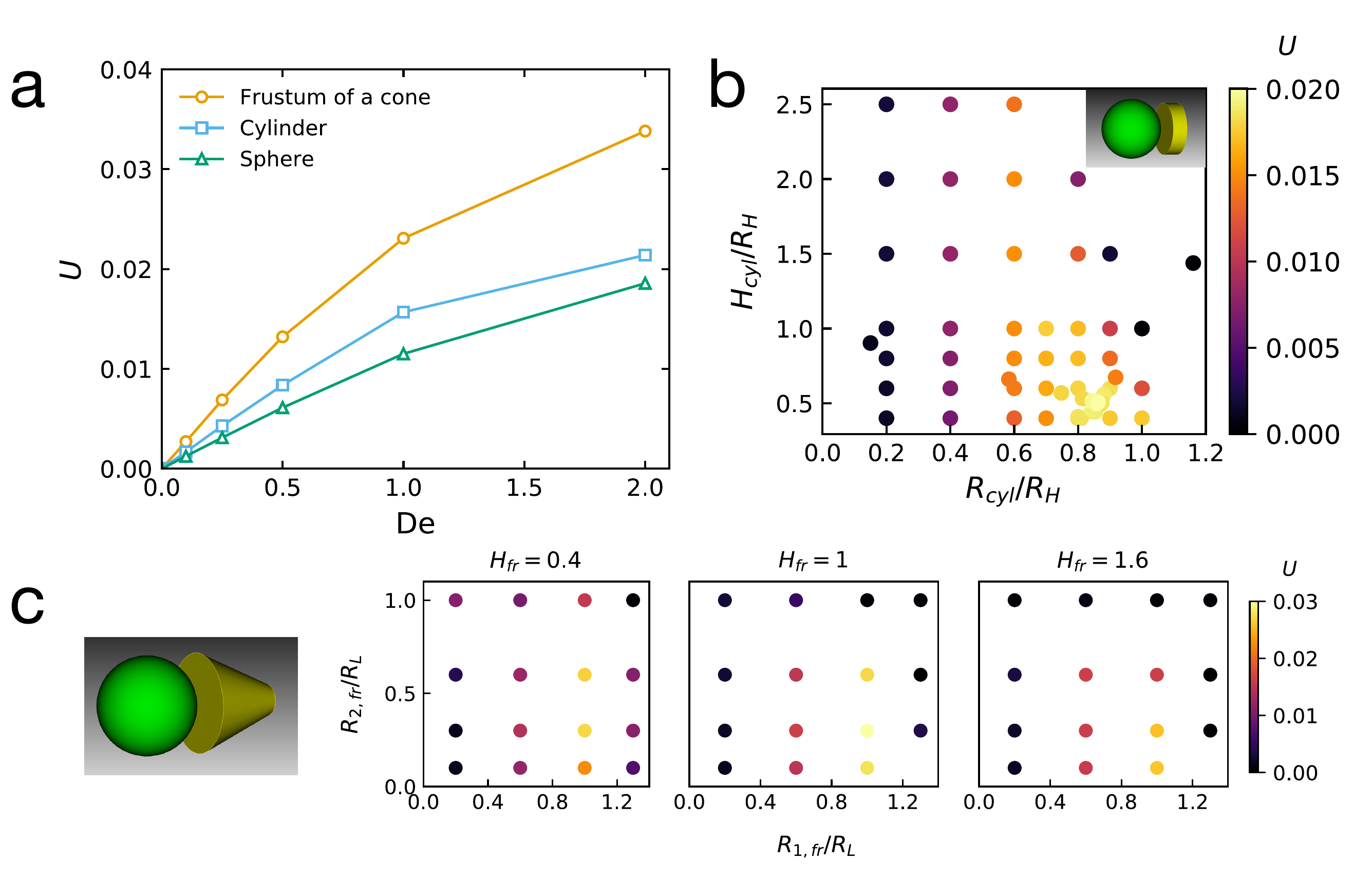}
    \caption{(a) Dimensionless swimming speed $U$ as a function of the Deborah number De for various tail shapes (as predicted by numerical simulations using the single-mode Giesekus equation with $\alpha_m=0.035$, $\hst=0.17$, $\beta=0.46$). For all De, the swimmer with optimal cylindrical tail ($R_{cyl}/R_L=0.86$ and $H_{cyl}/R_L=0.51$) is faster than that with the optimal spherical tail ($\rst=0.75$, as determined in prior work \citep{binagia2021self}). Likewise, the swimmer with the optimal conical frustum tail ($R_{2,fr}/R_L=1.0$, $R_{1,fr}/R_L=0.31$, $H_{fr}/R_L=1.1$) is faster than that having the cylindrical tail for all De. (b) The set of cylindrical tail shapes that were considered in finding the locally optimal tail used in subfigure (a). The inset depicts the geometry of the optimally-shaped cylindrical tail. (c) For the optimization study of the frustum, a 3-dimensional shape space must be explored. We show our results for 3 different cone height values, the forward speed as a function of the other two geometric parameters ($R_{1,fr}$ and $R_{2,fr}$), the radii of the circular cross-sections at each end of the tail (where $R_{1,fr}$ refers to the radius closest to the swimmer's head). Both the parameter sweeps performed in (b) and (c) were conducted at De = 1. }
    \label{fig:different-tail-shapes}
\end{figure}

Three distinct tail shapes were considered in this process: spherical tails, cylindrical tails, and conical tails (specifically cones with the tip of the cone removed, otherwise known as a frustum). The details of the optimization procedure used to ascertain the optimal geometry for each class of tail shapes is discussed in \cref{sec:tail-optimization}. Note that the optimal shape of the sphere (i.e. $\rst=0.75$) has been determined in prior work \citep{binagia2021self}. These locally optimal shapes where then used to determine the swimmer's speed as a function of De using the single-mode Giesekus equation with $\alpha_m=0.035$, as shown in \cref{fig:different-tail-shapes}. We see that for all De, the speed of the optimal cylindrical tail is strictly greater than the optimal spherical tail; likewise, the speed of the conical frustum is always greater than that of the cylindrical tail. This figure suggests that the swimming speed of the robot with an optimal spherical tail could be more than doubled by constructing a tail in the shape of a frustum. 

In \cref{fig:different-tail-shapes}b, we visualize the variety of cylindrical tail shapes that were considered in locating the optimal geometry. From this figure, it is clear that the speed for a swimmer with a cylindrical tail is much more sensitive to the radius of the cylinder, $R_{cyl}$, than to the height of the cylinder, $H_{cyl}$. One implication of this, in designing these types of robotic swimmers, is that one may select the optimal cylinder radius to maximize propulsion but then select the height of the cylinder (or for other shapes, secondary geometrical parameters) to maximize additional objectives for the swimmer, i.e. maximize efficiency or stability or to minimize volume or mass, etc. 
Likewise, in \cref{fig:different-tail-shapes}c, we visualize the simulations conducted for various conical frustum tail shapes. From this plot, we see that a relatively large frustum radius near the head is desirable (the optimal value of $R_{1,fr}/R_L$ is 1.0), while a smaller radius seems to be desirable near the rear of the swimmer (the optimal value of $R_{2,fr}/R_L$ being 0.31). In contrast to the dependence on the radii, the dependence on the length of the cone is much less substantial, with the optimal cone length $H_{fr}/R_L=1.1$.




\section*{Discussion}
\label{sec:discussion}


The observable behaviors of robots are typically programmed to achieve some specific task. Here we discuss a kind of inherently perceptive robot, where behavior is directly encoded by its physical surroundings, without the use of traditional sensors or active control. Mimicking attributes of micro-biological swimmers, we have built an untethered swimmer that passively adapts to propel itself forward at different speeds, depending on the properties of the surrounding fluid. In this way, the behavior itself acts passively as a sensor to the observer. This flipped paradigm is not a completely novel concept in robotics or in biophysics \citep{bull2021excitable} \citep{campas2014quantifying} \citep{aydin2019physics}, but it is often difficult to achieve using simple physical systems. The functionality of this system implies a clear application: a swimming rheometer. 

\begin{figure}[ht!]
    \centering
    \includegraphics[width=0.6\linewidth]{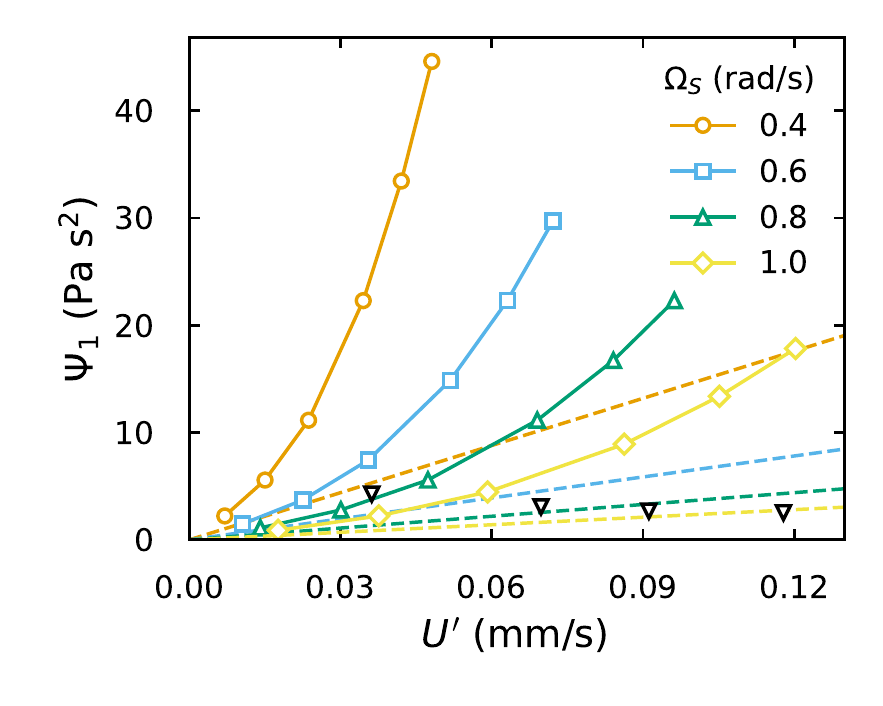}
    \caption{The value of the first normal stress coefficient $\Psi_1$ that the robot infers as it is observed to swim at (dimensional) speed $U'$ at a prescribed tail rotation rate $\Omega_S$. Here, we consider the viscosity to be known a-priori from independent measurements ($\mu_0=4.46$ Pa s). The dashed lines denote predictions from our low-De theory (i.e. \cref{eq:first-eq} with $\hst=0.17$, $\rst=0.43$, and assuming $\Psi_2\approx 0$). The solid lines and markers are predictions from our numerical simulations using the multi-mode Giesekus equation (i.e. the simulation data shown in \cref{fig:u-vs-de}, using the fact that $\Psi_1=2\sum \lambda^{(i)} \mu^{(i)}_p$ for a multi-mode model). Lastly, the black markers denote predictions for $\Psi_1$ based on our low De theory (\cref{eq:first-eq}) applied to the experimental data presented in \cref{fig:u-vs-de}c. This figure illustrates the operating characteristics of the swimming robot in terms of its use as a sensor. For example, operating at a small value of $\Omega_S$ (e.g. $\Omega_S=0.4$) allows the robot to be more sensitive to small changes in $U'$ but necessarily requires a greater experimental resolution of the observed swimming speed. We also observe that as $U'$ decreases, the accuracy of our low De theoretical prediction for $\Psi_1$ increases.}
    \label{fig:design_curve}
\end{figure}

\Cref{fig:design_curve} depicts a design curve, constructed from our simulation data and our theoretical results at low Deborah number. Observation of the propulsive speed of the robot (x-axis) directly implies a fluid property -- the first normal stress coefficient. This is exemplified by \cref{eq:first-eq}, where the rheology parameter we want to infer (i.e. $\Psi_1+2\Psi_2$) is written directly as a function of observables (the swimmer's speed and tail rotation rate) and known quantities (i.e. the swimmer's geometry, the viscosity of the fluid, etc.). Furthermore, very often it is the case in polymeric fluids that $|\Psi_2| \ll \Psi_1$ \citep{bird1987dynamics}, so upon assuming $\Psi_2 \approx 0$, \cref{eq:first-eq} yields a direct expression for $\Psi_1$ based upon measurements of our swimming robot. 

Alongside the prediction from our low De theory, we can also map out an "implicit" relationship between $\Psi_1$ and $U'$ defined by our numerical simulations using the multi-mode Giesekus equation (the simulation data presented in \cref{fig:u-vs-de}c). We use the fact that $\Psi_1=2\sum \lambda^{(i)} \mu^{(i)}_p$ for a multi-mode model to relate the relaxation spectrum $\lambda^{(i)}$ (i.e. the parameters input to the Giesekus equation) to $\Psi_1$, the relevant rheological property that would be measured at a benchtop rheometer.

From \cref{fig:design_curve}, we observe that while the low De asymptotic theory predicts linear design curves, the numerical simulations appear to be nonlinear with respect to the observed speed; this fact can be appreciated by noting that the curve of $\Psi_1$ vs. $U'$ is in many ways an inversion of \cref{fig:u-vs-de}c (i.e. now speed is on the x-axis and a measure of the fluid elasticity is on the y-axis). While the asymptotic theory is advantageous in that it provides a closed-form relationship between $\Psi_1$ and the swimmer's geometry, observed speed, etc., its usage should be confined to cases where the assumptions of its derivation hold (i.e. Re = 0, De $\ll 1$, an unconfined swimmer). In contrast, the curves mapped out by the numerical simulations are valid for finite De and can consider arbitrary levels of confinement; however, they only define an implicit relationship between $U'$ and $\Psi_1$ as shown in \cref{fig:design_curve}. One possible solution to circumvent this challenge is to conduct a finite set of numerical simulations (for a set rotation rate, $\Omega_s$), as done here, and then to use machine learning or other nonlinear curve fitting techniques to essentially enable interpolation between the points at which simulations were conducted. 

One consequence of \cref{fig:design_curve} is that different robot tail speed settings (different values of $\Omega_{S}$) should reproduce the same rheological property of the fluid, but with different sensitivities. This is related to the operating conditions: a rotation rate should be chosen such that both the sensitivity of the signal and the desired sensing range are taken into account. This design space may be constructed for other, more efficient tail shapes such as the frustum or a cylinder.

A corollary associated with this operating curve also holds: if maintaining a constant motor speed and the forward propulsion speed changes in time, this indicates a non-homogeneous fluid, where rheological properties of the fluid are changing spatially. This allows an observer to map non-uniform rheological properties within a region. 

To provide a proof of concept for our swimming rheometer, we can compare the inferred values of $\Psi_1$ based on our experimental observations and using \cref{eq:first-eq} to that which we would expect to get from a benchtop rheometer. The experimental data points displayed on \cref{fig:u-vs-de}c correspond to average tail rotation rates of 0.46, 0.73, 0.89, and 1.0 rad/s and swimming speeds of 0.036, 0.070, 0.091, and 0.12 mm/s. Plugging these values into \cref{eq:first-eq}, we would obtain respective predictions of $\Psi_1$ (assuming here again $\Psi_2\approx 0$) equal to 4.3, 3.1, 2.7, and 2.5 Pa $\mathrm{s}^2$ respectively (shown as the black markers on \cref{fig:design_curve}). It should be emphasized at this point that typical commercial rheometers can not measure $\Psi_1$ at these low shear rates \citep{morrison2001understanding}; this is clearly seen in \cref{fig:rheology}c, where $\Psi_1$ can only be reliably measured above a shear rate of 10 $s^{-1}$ given the sensitivity of the rheometer's normal force transducer (according to the ARES-G2 rheometer specification sheet, the minimum normal force that can be resolved is 0.001 N). For this reason, we compare our inferred values of $\Psi_1$ from our swimming rheometer to that predicted by our multi-mode Giesekus fit (i.e. the red dashed line in \cref{fig:rheology}c) at shear rates corresponding to the average shear rate created by a sphere rotating at rotation rate $\Omega_S$ (where the average shear rate for a sphere of radius $R$ rotating at rotation rate $\Omega$ at $\mathrm{Re}=0$ is equal to $\frac{\int_0^\pi|\frac{du_\phi}{dr}|_{r=R}|d\theta}{\int_0^\pi d\theta}=4\Omega/\pi$).

At these shear rates, the fit to the multi-mode Giesekus equation are 3.1, 2.2, 1.8, and 1.7 Pa $\mathrm{s}^2$. If we take the values predicted by the Giesekus fit as the "ground truth", we obtain percent differences (as compared to values using our swimming robot) of 38\%, 43\%, 47\%, and 50\% respectively. Because $\Psi_1$ is a dimensional quantity that can range across many orders of magnitude \cite{morrison2001understanding} \cite{bird1987dynamics}, it is also relevant to report that the estimated absolute error on $\Psi_1$ is computed to be $0.8-1.2$ Pa s$^2$ at these low shear rates (under 1 hz).

 This comparison between our inferred values of $\Psi_1$ from experiments and those predicted from a fit to $\Psi_1$ measurements taken using a commercial rheometer are in reasonable agreement, considering the assumptions of our asymptotic theory used to derive \cref{eq:first-eq}. It is notable that the swimming rheometer is effectively providing a direct estimate for $\Psi_1$ at shear rates where commercial rheometers typically cannot. Lastly, we remark that as the tail rotation rate decreases, i.e. decreasing De, the accuracy of our low De theory and formula for $\Psi_1$ increases; indeed, this is clearly manifested above in that we observe the relative percent errors decreasing with decreasing $\Omega_S$. As remarked in the discussion around \cref{fig:design_curve}, an even more accurate prediction for $\Psi_1$ at the present experimental De values would involve interpolating between the discrete simulation data presented in \cref{fig:design_curve} so as to determine an implicit mapping from $U'$ and $\Omega_S$ to $\Psi_1$. 

Additionally, it is notable that both the first and second normal stress coefficients, $\Psi_1$ and $\Psi_2$, can be measured simultaneously if one makes an independent measurement of the head rotation rate of the swimmer. This fact can be appreciated by rewriting our asymptotic equation for the head rotation rate, \cref{eq:ang-vel-formula}, directly in terms of $\Psi_1$ and $\Psi_2$:
\begin{equation}
    \bar{\Omega}_L=\rst^3\frac{\bigg( 1-\frac{1}{(1+\rst+\hst)^3}
    -\frac{2(\Psi_1 \Omega_S/(2\mu_0))^2}{15}(1-2\Psi_2/\Psi_1)^2\bigg)}{1-\frac{\rst^3}{(1+\rst+\hst)^3}}
    \label{eq:ang-vel-formula-text}
\end{equation}
That is, just as was the case for \cref{eq:first-eq} and the swimming speed, we now have a direct relationship between the head rotation rate and rheological parameters of the fluid. This means that if, for example, the viscosity of a fluid is known, \cref{eq:first-eq} and \cref{eq:ang-vel-formula-third-order} can be solved simultaneously as a system of two equations for the variables $\Psi_1$ and $\Psi_2$ after measuring the swimming speed and head rotation of our swimmer for a prescribed tail rotation rate. Alternatively, if one is to assume $\Psi_2\approx 0$ (which is often a reasonable estimate for polymeric fluids), one can use these two coupled equations to solve for $\Psi_1$ and the viscosity $\mu_0$. We point out that the potential to measure both $\Psi_1$ and $\Psi_2$ is especially exciting in the field of rheology where traditionally measurements of $\Psi_2$ have proved quite challenging, requiring both fine instrument sensitivity (since typically $|\Psi_2|\ll \Psi_1$) as well as unique instrument configurations \citep{morrison2001understanding}; for example, to measure $\Psi_2$ using a cone-and-plate rheometer requires outfitting the rheometer with a series of pressure transducers along the plate (see for example \citep{baek2003monolithic}). Furthermore, recently proposed methods for measuring the normal stress coefficients, e.g. measuring the speed of a rotating dumbbell \citep{pak2012micropropulsion} or the relative force between two microrheological probes \citep{khair2010active}, require that two independent measurements be made to compute both $\Psi_1$ and $\Psi_2$; in contrast, both $\Psi_1$ and $\Psi_2$ can be computed from a single experiment in our methodology since we can observe simultaneously two measurable quantities (speed and head rotation rate) that each depend on the rheological parameters of the fluid in a distinctive, independent manner (i.e. comparing \cref{eq:first-eq} and \cref{eq:ang-vel-formula-text}).

Further work is ongoing to expand the number of measurable rheological properties. With additional on-board instrumentation, full characterization of the loss modulus ($G''$) and the viscosity are also theoretically possible, by examining transient responses of the robot. These effects are much easier to observe when the geometry of the robot has been fully optimized to maximize the observable signals. 

The passive sensing aspect of this robot (and therefore its relative simplicity) opens exciting opportunities for extreme miniaturization. We have developed several designs for far simpler controllers (specifically, a purely analog controller to drive DC motors via an external optical PWM signal) that could be easily incorporated into robots 1-2 orders of magnitude smaller in size. The eventual miniaturization of such devices on the scale of 100-500 $\mu$m using MEMS techniques could open numerous opportunities for applications in microbiology, environmental sensing and medical diagnostics.   


\section*{Methods} 
\label{sec:methods}


\subsection{Mathematical Description of the Swimming Problem}
\label{sec:math}


The fluid motion is governed by conservation of momentum and mass, written in dimensionless form as: 
\begin{equation}
    \mathrm{Re}\bigg(\frac{\partial\mathbf{u}}{\partial t}+\mathbf{u}\cdot\nabla\mathbf{u}\bigg)= \nabla\cdot
    \bm{\sigma}, \quad\quad \nabla\cdot\mathbf{u}=\mathbf{0}
    \label{nd_stokes}
\end{equation}
where $\mathbf{u}$ is the velocity of the fluid, $t$ is time, and $\bm{\sigma}$ is the Cauchy stress. These equations were made dimensionless by scaling lengths by the radius of the large sphere $R_L$, time by the inverse of the smaller sphere's rotation rate, $\Omega_S^{-1}$, velocities by the product $R_L \Omega_S$, and stresses by $\mu_0 R_L^2 \Omega_S$ where $\mu_0$ is the total zero-shear viscosity of the fluid. Note that we choose to use $\Omega_S$ rather than $\Omega_L$ in making time dimensionless because the former is prescribed in our model and experiments, while the latter is a dependent variable that is observed. With this choice of characteristic scales, the Reynolds number Re is defined as $\mathrm{Re}=\rho R_L^2 \Omega_S/\mu_0$ where $\rho$ is the density of the fluid. We assume $\mathrm{Re}=0$ for the analytical theory and numerical simulations presented in this work since we are generally interested in modeling the motion of microorganisms, which swim at virtually zero Reynolds number \citep{purcell1977life}, and because the viscosity of our experimental fluid is sufficiently large to warrant this assumption. 

The fluid in which the swimmer is immersed is viscoelastic; consequently, we write the total stress $\bm{\sigma}$ as the sum of a Newtonian and polymeric contribution: 
\begin{equation}
    \bm{\sigma} = -p\mathrm{\mathbf{I}} + \beta(\nabla\mathbf{u} + \nabla\mathbf{u}^T) + \bm{\tau}^p
    \label{eq:total-stress}
\end{equation}
where $p$ denotes pressure, $\bm{\tau}^p$ is the extra stress coming from deformation of polymer molecules immersed in the fluid, and $\beta=\mu_s/(\mu_s+\mu_p)=\mu_s/\mu_0$ is the ratio of the solvent viscosity $\mu_s$ to the total zero-shear solution viscosity $\mu_0$ (for a fluid with polymer viscosity equal to $\mu_p$). In general, we use the Giesekus constitutive equation to describe the extra polymer stress $\bm{\tau}^p$:
\begin{align}
&\bm{\tau}^{p} = \frac{1-\beta}{\text{De}}(\mathbf{c}-\mathbf{I}) \label{eq:giesekus-1} \\
&\text{De} \stackrel{\triangledown}{\mathbf{c}} + (\mathbf{c}-\mathbf{I}) + \alpha_m(\mathbf{c}-\mathbf{I})^2   = \mathbf{0}.
\label{eq:giesekus}
\end{align}
In the above equation, $\mathbf{c}$ is the conformation tensor and  $\stackrel{\triangledown}{\mathbf{c}}=\partial \mathbf{c}/\partial t+\mathbf{u}\cdot\nabla\mathbf{c}-\nabla\mathbf{u}^T\cdot\mathbf{c}-\mathbf{c}\cdot\nabla\mathbf{u}$ is the upper-convected derivative. $\mathrm{De}=\lambda \Omega_S$ is the Deborah number, which describes the relative importance of elastic effects in a viscoelastic fluid with relaxation time $\lambda$. 
If the Giesekus mobility parameter $\alpha_m$ is chosen to be equal to zero, the simpler Oldroyd-B model is recovered, which models the polymer molecules in the fluid to be Hookean dumbbells \citep{oldroyd1950formulation}. Nonzero values of $\alpha_m$ allow the Giesekus equation to model drag anisotropy experienced by the Hookean dumbbells; this permits the model to predict more realistic rheological behavior such as shear-thinning and nonzero second normal stress differences \citep{bird1987dynamics}. 

\Cref{eq:giesekus-1,eq:giesekus} describes a fluid with a single relaxation time; our model can easily be extended to consider multiple relaxation times by considering a multi-mode model for the fluid, where the total extra polymer stress is simply the sum of that contributed by each mode: 
\begin{equation}
    \bm{\tau}^{p} = \sum_{i=0}^n \bm{\tau}^{p}_i.
\end{equation}
for a model with $n$ modes. In this formulation, each mode has its own polymer viscosity $\mu_p^{(i)}$, relaxation time $\lambda^{(i)}$, and Giesekus mobility parameter $\alpha_m^{(i)}$ and the stress for each individual mode is $\bm{\tau}^{p}_i$ is described by \cref{eq:giesekus-1,eq:giesekus}. It is convenient to define an "average" relaxation time for the multi-mode fluid, defined as follows:
\begin{equation}
    \bar{\lambda}=\frac{\sum_{i=0}^n \lambda^{(i)} \mu_p^{(i)}}{\sum_{i=0}^n \mu_p^{(i)}}
\end{equation}

To close our mathematical model for the motion of the swimming robot, we must provide a set of appropriate boundary conditions and constraints applied to the swimming motion. The no-slip boundary condition applied at the surface of the swimmer gives $\mathbf{u}_S=\mathbf{U}+\mathbf{\Omega}_S\times \mathbf{r}_S$ for the velocity at the surface of the small sphere and $\mathbf{u}_L=\mathbf{U}+\mathbf{\Omega}_L\times \mathbf{r}_L$ for the velocity at the surface of the large sphere. $\mathbf{U}$ denotes the translational velocity of the swimmer and $\mathbf{r}_S$ and $\mathbf{r}_L$ are position vectors originating from the center of the smaller and larger sphere respectively. Finally, applying conservation of linear and angular momentum to the swimming body at $\mathrm{Re}=0$ gives the well-known result that the net force and torque on the swimmer must each be equal to zero \citep{elfring2015theory}: 
\begin{align}
    &\mathbf{F}=\int_S\bm{\sigma}\cdot\mathbf{n}dS = \mathbf{0},\\
    &\mathbf{L}=\int_S \mathbf{r}\times(\bm{\sigma}\cdot\mathbf{n})dS=\mathbf{0}.
\end{align}
where $S$ denotes the surface of the body, $\mathbf{r}$ is a position vector originating from the center of the gap between the two spheres, and $\mathbf{n}$ is the unit normal to the surface of the swimmer. Lastly, in the lab frame of references, the fluid far away from the swimmer is quiescent so that $\mathbf{u}(|\mathbf{r}|\rightarrow \infty)=0$.

\subsection{Small De Asymptotic Theory for the Swimming Speed}
\label{sec:theory-section}

For sufficiently small De, the swimming speed can be predicted via an asymptotic theory developed by \citet{binagia2021self}. While we refer the interested reader to \citep{binagia2021self} for the details of the derivation, we will briefly summarize the main ideas and results for this analytical theory since it is used frequently in the present work. The main result is a prediction for the dimensionless velocity of the swimmer $U_z$:
\begin{equation}
    U_z = U_z^0+U_z^{(HO)}
    \label{eqn:uz-split}
\end{equation}
where 
\begin{equation}
    U_z^{0} = \text{De}(1-\beta)\frac{(1+\hst)^2\rst^3(1+\hst+3\rst)\ost^2-\rst(\hst+\rst)^2(3+\hst+\rst)}{(1+\rst)(1+\hst+\rst)^5\ost^2}
    \label{eqn:uz0}
\end{equation}
and
\begin{equation}
    U_z^{(HO)} = \text{De}(1-\beta)\rst^3\bigg(\frac{Q}{3(1+\rst)(1+\hst+\rst)^7\ost^2}\bigg).
    \label{eqn:uzHO}
\end{equation}
$Q$, the factor in the numerator of \cref{eqn:uzHO} is given by: 
\begin{eqnarray} 
    Q(\rst,\hst,\Omega^*)=-3+15\hst+6\hst^2+2\hst^3+15\rst+12\hst\rst+6\hst^2\rst+6\rst^2+6\hst\rst^2\nonumber    \\
    +2\rst^3-\big[2(1+\hst)^3+6(1+\hst)^2\rst+15(1+\hst)\rst^2-3\rst^3\big]\ost^2.
\end{eqnarray}
We use the subscript $z$ to denote the fact that the velocity is in the vertical $z$ direction, i.e. along the axis of rotation (c.f. \cref{fig:fig-1-schematic}b). These equations are written in terms of several key dimensionless numbers, including the ratio of the size of the two spheres $\rst=R_S/R_L$, the dimensionless separation between the two spheres $\hst=h_{sep}/R_L$, and the ratio of the spheres' rotation rates $\ost=\Omega_S/\Omega_L$. As discussed in \citep{binagia2021self}, this expression for the swimming speed is derived by considering far-field hydrodynamic interactions between two spheres rotating in isolation at $\mathrm{Re}=0$. The flow field for sphere is assumed to be the leading order viscoelastic flow for a rotating sphere in Stokes flow \citep{bird1987dynamics}. Furthermore, since we are considering a weakly elastic flow, we assume the hydrodynamic drag for each sphere is simply that given by Stokes drag law \citep{kim2013microhydrodynamics}. To enforce the fact that the swimmer should be torque free, we set $\Omega^*=(1/\rst)^3$, a relationship that is derived by equating the dimensional torque each sphere experiences when rotating at $\mathrm{Re}=0$ in a Newtonian fluid, i.e. $8\pi\mu_0 \Omega R^3$. Despite these simplifying assumptions, it is shown in \citep{binagia2021self} that this simple analytical theory agrees with numerical simulations up to $\mathrm{De} \approx 1$ for a wide range of values for the viscosity ratio $\beta$; this is likely a consequence of the fact that the disturbance flow created by the rotation of each sphere decays relatively rapidly for problems of this kind, decaying as $1/r^2$ where $r$ is the distance from the center of each sphere. We finally remark that for all of the sets of parameters considered in \citep{binagia2021self} as well as in the present work, $U_z>0$, corresponding to propulsion in the direction of the large sphere (c.f. \cref{fig:fig-1-schematic}b).

The above set of equations were derived in the context of the Oldroyd-B constitutive equation (i.e. \cref{eq:giesekus,eq:giesekus-1} with $\alpha_m=0$) and are thus written in terms of the rheological parameters $\mu_p$ and $\lambda$; however, it is quite straightforward to express these results in terms of another common constitutive equation, the second-order fluid (SOF) model. The SOF model, which applies to slowly varying flows, is formally derived by applying the retarded motion expansion to a given flow and retaining terms quadratic in the velocity gradient \citep{bird1987dynamics}. The deviatoric stress (i.e. $\bm{\tau}=\beta(\nabla\mathbf{u} + \nabla\mathbf{u}^T) + \bm{\tau}^p$) for the SOF model is given by:
\begin{equation}
    \bm{\tau}=\bm{\dot{\gamma}}-\mathrm{De_{SO}}(\stackrel{\triangledown}{\bm{\dot{\gamma}}}+B \bm{\dot{\gamma}}\cdot \bm{\dot{\gamma}})
    \label{eq:sof}
\end{equation}
where $\bm{\dot{\gamma}}=\nabla\mathbf{u} + \nabla\mathbf{u}^T$, $\mathrm{De_{SO}}=(\Psi_1 \Omega_S)/(2\mu_0)$ is the Deborah number defined for the SOF model, $B=-2\Psi_2/\Psi_1$, and $\Psi_1$ and $\Psi_2$ are the first and second normal stress coefficients respectively. Note that because $\lambda=\frac{\Psi_1}{2\mu_p}$ , $\mathrm{De_{SO}}=\mathrm{De}(1-\beta)$ \citep{bird1987dynamics}. From \cref{eq:sof}, it is clear that the second-order fluid model captures the first deviations from Newtonian behavior (where simply $\bm{\tau}=\bm{\dot{\gamma}}$). 

Binagia and Shaqfeh \citep{binagia2021self} showed that the swimming speed for the two-sphere swimmer is proportional to the radial disturbance flows created by each rotating sphere. The leading-order radial flow created by a rotating sphere in a viscoelastic fluid is identical for the Oldroyd-B and second-order fluid models; written dimensionlessly in terms of the parameters of the latter, it is: $u_r=\mathrm{De_{SO}}(1-B)[(1/2r^2-3/2r^4+1/r^5)(3\cos^2\theta-1)$ \citep{bird1987dynamics}. We can write this in terms of the Oldroyd-B fluid by noting that $\Psi_2=0$ (and hence $B=0$) for the Oldroyd-B fluid and $\mathrm{De_{SO}}=\mathrm{De}(1-\beta)$: $u_r=\mathrm{De}(1-\beta)[(1/2r^2-3/2r^4+1/r^5)(3\cos^2\theta-1)$. \citet{binagia2021self} presented this final result, linear in $\mathrm{De}(1-\beta)$ (since the swimming speed is proportional to the radial flow velocity in this asymptotic theory), but we will now allow for nonzero $\Psi_2$, such that in general the swimming speed takes the form: 
\begin{equation}
    U = |U_z| = \mathrm{De_{SO}}(1-B)f(\rst,\hst,\ost)
\end{equation}
where $f(\rst,\hst,\ost)$ captures the complex dependence on the geometry and kinematics of the swimmer (i.e. as shown in \cref{eqn:uz-split,eqn:uz0,eqn:uzHO}). Writing the expression for speed in this way, in terms of $\Psi_1$ and $\Psi_2$, allows us to explicitly rearrange for these direct measures of the rheology of the fluid:
\begin{align}
    \frac{U'}{\Omega_S R_L} &= \bigg(\frac{\Psi_1 \Omega_S}{2 \mu_0}\bigg)(1+2\Psi_2/\Psi_1)(f(\rst,\hst,\ost) \\
    \Psi_1+2\Psi_2 &=\frac{2 U' \mu_0}{\Omega_S^2 R_L f(\rst,\hst,\ost)}.
    \label{eqn:psi1-eq}
\end{align}
\Cref{eqn:psi1-eq} illustrates how rheological properties of the fluid ($\Psi_1$ and $\Psi_2$) can be directly inferred from a measurement of the dimensional swimming speed $U'$ of the robot for a given tail rotation rate $\Omega_S$. It also illustrates how the geometry of the swimmer, i.e. the function  $f(\rst,\hst,\ost)$, directly affects the mapping between the measured speed $U'$ and the inferred properties $\Psi_1$ and $\Psi_2$; in other words, the geometry should be carefully selected to maximize the utility of the device as a sensor for the fluid. We note at this point that, unless one assumes $\Psi_2\approx 0$, $\Psi_1$ and $\Psi_2$ cannot both be determined via \cref{eqn:psi1-eq}; rather, we can only compute the sum $\Psi_1+2\Psi_2$. As we will soon discuss in the following section, however, an independent measurement of the head rotation rate of the swimmer provides an additional equation that, when solved alongside $\cref{eqn:psi1-eq}$, allows one to determine uniquely both the values of $\Psi_1$ and $\Psi_2$. 

Lastly, while the expression above was derived in the context of a single-mode fluid, the extension to a multi-mode model is straightforward. The effect of the multiple modes can be shown to be additive, such that their impact on the swimming speed is completely captured by the mean of their individual effects. Put more precisely, the swimming speed for the swimmer immersed in a multi-mode fluid is given by \cref{eqn:uz-split,eqn:uz0,eqn:uzHO} but with the Deborah number De now defined in terms of the \textit{average} relaxation time of the fluid $\mathrm{De}=\bar{\lambda}\Omega_S$. This allows us to make direct comparisons between experimental data, where the fluid is fit to a multi-mode constitutive equation, with simulations and theory for a single-mode fluid.

\subsection{Small De Asymptotic Theory for the Head Rotation Rate}
An expression for the rotation rate of the head (i.e. the larger sphere) can be obtained by enforcing the torque-free condition on the swimmer. In the case of two spheres rotating in a Newtonian fluid at $\mathrm{Re}=0$ and neglecting hydrodynamic interactions, the balance is simply:
\begin{equation}
    -8\pi \mu_0 R_L^3 \Omega_L+8\pi \mu_0 R_S^3 \Omega_S=0
\end{equation}
where again $\Omega_L$ and $\Omega_S$ denote the \textit{magnitude} of the rotation rates. Thus in a Newtonian fluid at Stokes flow, the ratio of the rotation rates $\Omega_L/\Omega_S=(R_S/R_L)^3=\rst^3$. Of course, there will be a modification to the torque experienced by each sphere when rotating in a viscoelastic fluid; Walters and Waters have performed a perturbation expansion for a sphere rotating in a third-order fluid and found that the hydrodynamic torque $L'$ experienced by the sphere is given by \citep{walters1963use,walters1964interpretation,walters1964steady}:
\begin{equation}
    L'=-8\pi \mu_0 R^3 \Omega\bigg[ 1 + \frac{\mathrm{Re}^2}{1200}+\frac{\mathrm{Re}\mathrm{De_{SO}}}{140}\bigg(1+\frac{b_{11}}{b_2}\bigg)-\frac{2\mathrm{De_{SO}}^2}{15}\bigg(1+\frac{b_{11}}{b_2}\bigg)^2-\frac{24\mathrm{De_{SO}}^2}{5}\frac{b_1(b_{12}-b_{1:11})}{b_2^2}\bigg]
\end{equation}
where $b_1$, $b_{12}$, $b_{1:11}$, and $b_{2}$ are parameters of the third-order fluid model \citep{bird1987dynamics}. Here, $\mathrm{Re}=\frac{\rho R^2 \Omega}{\mu_0}$ and $\mathrm{De_{SO}}=(\Psi_1\Omega)/(2\mu_0)=-b_2\Omega/b_1$. For $\mathrm{Re}=0$, we can rewrite this as:
\begin{equation}
    L'=-8\pi \mu_0 R^3 \Omega\bigg[ 1-\frac{2\mathrm{De_{SO}}^2}{15}(1+B)^2-\frac{24\mathrm{De_{SO}}^2}{5}(B_{12}-B_{1:11})\bigg]
\end{equation}
where $B=b_{11}/b_2=-2\Psi_2/\Psi_1$, $B_{12}=b_{12}b_1/b_2^2$, and $B_{1:11}=b_{1:11}b_1/b_2^2$. From this equation, it's readily apparent that as the elasticity of the fluid increases (increasing De), the hydrodynamic torque experienced by the rotating sphere decreases. In summary, each sphere experiences a Newtonian torque $L_{newt}'=8\pi \mu_0 R^3 \Omega$ and an additional torque from rotating in an elastic fluid, $L_{elastic}'$, equal to: 
\begin{equation}
    L_{elastic}'=8\pi \mu_0 R^3 \Omega\bigg[ \frac{2\mathrm{De_{SO}}^2}{15}(1+B)^2+\frac{24\mathrm{De_{SO}}^2}{5}(B_{12}-B_{1:11})\bigg]
\end{equation}

The final contribution we would like to consider in our torque balance is that from hydrodynamic interactions. Note that for this, we need only consider hydrodynamic interactions given by the Newtonian disturbance flow created by each sphere; this is because to leading-order, fluid elasticity has zero effect on the azimuthal component of velocity for a rotating sphere \citep{bird1987dynamics}. Indeed, given the alignment of our spheres in the vertical direction, the only component of velocity that can alter the torque experienced by the other sphere is the azimuthal component. Applying Faxen's law for the hydrodynamic torque applied to one sphere as a result of the disturbance flow from the other sphere \citep{kim2013microhydrodynamics}, we obtain the following contributions to the torque on each half of the swimmer's body:
\begin{align}
    L_{HI,L}'&=-8\pi \mu_0 R_L^3 \bigg( \frac{R_S^3 \Omega_S}{(R_S+R_L+h_{sep})^3}\bigg) \\
    L_{HI,S}'&=8\pi \mu_0 R_S^3 \bigg( \frac{R_L^3 \Omega_L}{(R_S+R_L+h_{sep})^3}\bigg)
\end{align}
Note that interactions between the two spheres reduce the net hydrodynamic torque that either of them experiences. A negative sign appears in the expression for $L_{HI,L}'$ given the convention we have taken in \cref{fig:fig-1-schematic}b where the larger sphere is seen to have a positive angular velocity with respect to the vertical $z$-axis (and thus the torque applied by the smaller sphere will act in the negative z-direction, in alignment with its negative angular velocity). 

\begin{figure}
    \centering
    \includegraphics{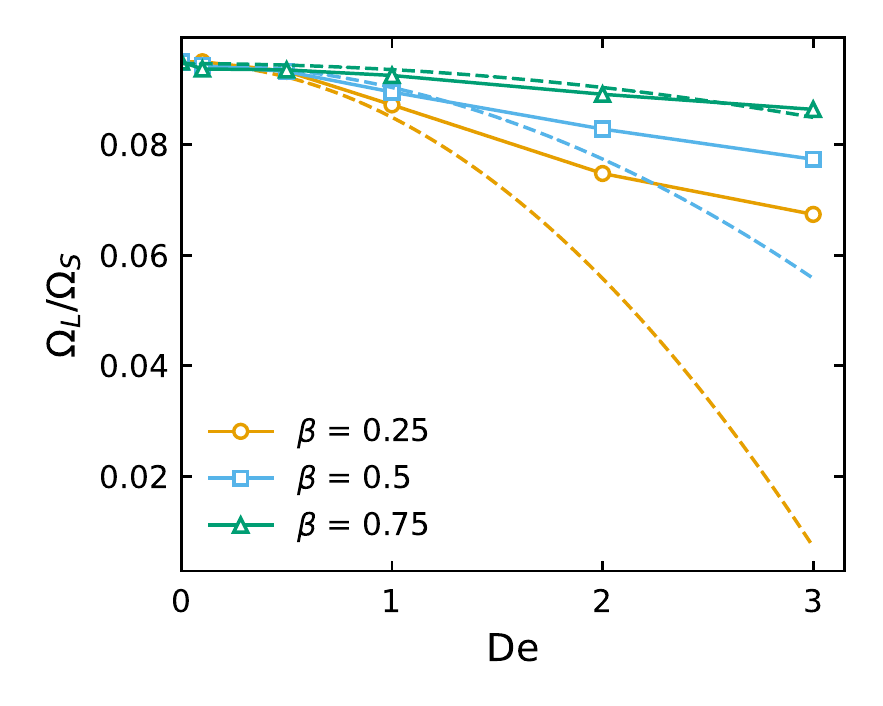}
    \caption{Rotation rate of the head of the swimmer ($\Omega_L$) normalized by that of the tail ($\Omega_S$) as a function of Deborah number (De) and the viscosity ratio ($\beta$). The markers denote numerical simulations while the dashed lines denote the asymptotic theory, i.e. \cref{eq:ang-vel-formula}; both consider $\hst=0.05$ and $\rst=0.5$.}
    \label{fig:ang-vel-validation}
\end{figure}

We can combine each of the above contributions into a total torque balance on the swimmer:
\begin{align}
    &L_L' + L_S' = 0\\
    &L_{newt,L}'+L_{elastic,L}'+L_{HI,L}'+L_{newt,S}'+L_{elastic,S}'+L_{HI,S}'=0 
\end{align}
\begin{multline}
-8\pi \mu_0 R_L^3 \Omega_L \bigg[1+ \frac{R_S^3 (\Omega_S/\Omega_L)}{(R_S+R_L+h_{sep})^3}-\frac{2\mathrm{De_{SO}}^2}{15}\bigg(\frac{\Omega_L}{\Omega_S}\bigg)^2(1+B)^2-\frac{24\mathrm{De_{SO}}^2}{5}\bigg(\frac{\Omega_L}{\Omega_S}\bigg)^2(B_{12}-B_{1:11})\bigg]\\
+8\pi \mu_0 R_S^3 \Omega_S\bigg[ 1+\frac{R_L^3 (\Omega_L/\Omega_S)}{(R_S+R_L+h_{sep})^3}
    -\frac{2\mathrm{De_{SO}}^2}{15}(1+B)^2-\frac{24\mathrm{De_{SO}}^2}{5}(B_{12}-B_{1:11})\bigg]=0
\end{multline}
In dimensionless form:
\begin{multline}
-\bar{\Omega}_L\bigg[1+ \frac{\rst^3 \bar{\Omega}_L^{-1}}{(1+\rst+\hst)^3}-\frac{2\mathrm{De_{SO}}^2}{15}\bar{\Omega}_L^2(1+B)^2-\frac{24\mathrm{De_{SO}}^2}{5}\bar{\Omega}_L^2(B_{12}-B_{1:11})\bigg]\\
+\rst^3\bigg[ 1+\frac{\bar{\Omega}_L}{(1+\rst+\hst)^3}
    -\frac{2\mathrm{De_{SO}}^2}{15}(1+B)^2-\frac{24\mathrm{De_{SO}}^2}{5}(B_{12}-B_{1:11})\bigg]=0
\end{multline}
where we have defined a dimensionless head rotation rate $\bar{\Omega}_L=\Omega_L/\Omega_S$. Here $\mathrm{De_{SO}}=(\Psi_1\Omega_S)/(2\mu_0)$ is defined using the rotation rate of the small sphere (the "tail" of the swimmer) as is done in the rest of the paper. At this point, we can simplify the torque balance by noting the following. We know from the torque balance in a Newtonian fluid that $\bar{\Omega}_L \sim \rst^3$. Thus, the elastic contribution to the torque on the small sphere is of size $O(\mathrm{De_{SO}}^2)$ while that on the large sphere is $O(\rst^9 \mathrm{De_{SO}}^2)$. Since $\rst \le 1$, clearly the latter will be significantly smaller than the former and can safely be neglected. That leaves us with the following expression:
\begin{multline}
-\bar{\Omega}_L\bigg[1+ \frac{\rst^3 \bar{\Omega}_L^{-1}}{(1+\rst+\hst)^3}\bigg]
+\rst^3\bigg[ 1+\frac{\bar{\Omega}_L}{(1+\rst+\hst)^3}
    -\frac{2\mathrm{De_{SO}}^2}{15}(1+B)^2-\frac{24\mathrm{De_{SO}}^2}{5}(B_{12}-B_{1:11})\bigg]=0
\end{multline}

\begin{multline}
\rst^3\bigg[ 1
    -\frac{2\mathrm{De_{SO}}^2}{15}(1+B)^2-\frac{24\mathrm{De_{SO}}^2}{5}(B_{12}-B_{1:11})-\frac{1}{(1+\rst+\hst)^3}\bigg]=\bar{\Omega}_L(1-\frac{\rst^3}{(1+\rst+\hst)^3})
\end{multline}
Solving for $\bar{\Omega}_L$:
\begin{equation}
    \bar{\Omega}_L=\rst^3\frac{\bigg( 1-\frac{1}{(1+\rst+\hst)^3}
    -\frac{2\mathrm{De_{SO}}^2}{15}(1+B)^2-\frac{24\mathrm{De_{SO}}^2}{5}(B_{12}-B_{1:11})\bigg)}{1-\frac{\rst^3}{(1+\rst+\hst)^3}}
    \label{eq:ang-vel-formula-third-order}
\end{equation}
This result shows that, in contrast to the swimming speed, fluid elasticity has no effect on the rotation rate of the large sphere to leading-order (i.e. at O(De)). Rather, the effect of elasticity manifests at $O(\mathrm{De}^2)$ and leads to a reduction in the dimensionless rotation rate of the large sphere. Note that if we taken the limit of $\hst\rightarrow \infty$ and $\mathrm{De_{SO}}=0$, we can confirm that we obtain $\bar{\Omega}_L=\rst^3$ (which is what we expect in the case of a Newtonian fluid and zero hydrodynamic interactions). This equation can also be simplified if we wish to omit the third-order fluid parameters by setting $B_{12}=0$ and $B_{1:11}=0$ (i.e. in effect only considering the second-order fluid model):
\begin{equation}
    \bar{\Omega}_L=\rst^3\frac{\bigg( 1-\frac{1}{(1+\rst+\hst)^3}
    -\frac{2\mathrm{De_{SO}}^2}{15}(1+B)^2\bigg)}{1-\frac{\rst^3}{(1+\rst+\hst)^3}}
    \label{eq:ang-vel-formula}
\end{equation}
This form of the equation is useful when we consider the application of our swimmer as a rheometer for the fluid since now the only rheological parameters appearing on the right-hand side are $\Psi_1$ and $\Psi_2$ via the definitions of $\mathrm{De_{SO}}$ and $B$ respectively. We can confirm the formula presented in \cref{eq:ang-vel-formula} by comparing it to numerical simulations using the single-mode Oldroyd-B equation, as shown in \cref{fig:ang-vel-validation}. To compare \cref{fig:ang-vel-validation} to numerical simulations using the Oldroyd-B equation, we note that for the Oldroyd-B equation $\Psi_2=0$ and thus $B=0$, and furthermore that the Deborah number defined for the Oldroyd-B equation, i.e. $\mathrm{De}=\lambda \Omega_S$, is related to that for the second- and third-order fluid models via $\mathrm{De_{SO}}=(1-\beta)\mathrm{De}$. The agreement between the theory and simulations suggests that our asymptotic equation is indeed an accurate measure of the head rotation rate for small De. 

\renewcommand{\figurename}{Supplementary Figure}

\subsection{Numerical Simulations}
\label{sec:simulations}

For the numerical calculations, the methodology largely follows that described in  \citep{binagia2021self} and is briefly summarized below. 3D simulations of the governing equations and boundary conditions are conducted through the use of a third-order accurate finite volume flow solver developed at Stanford's Center for Turbulence Research \citep{Ham2006}. This code has been thoroughly tested for accuracy and robustness for a wide range of problems, including viscoelastic flows \citep{Richter2010,padhy2013simulations,yang2016numerical}, deformable particles \citep{saadat2018immersed}, and active swimmers \citep{binagia2019three,binagia2020swimming,housiadas_binagia_shaqfeh_2021}. The problem is solved in the co-moving frame of reference, wherein the swimmer is stationary in a uniform flow, such that we may use a body-fitted mesh for our computational domain. The evolution equation for the conformation tensor (\cref{eq:giesekus}) is solved using the log-conformation method \citep{Fattal2004,Hulsen2005}, and for boundary conditions for the conformation tensor we set $\mathbf{c}=\mathbf{I}$ at the entrance to the domain and utilize a convective outlet boundary condition at the exit. We use a cylindrical computational domain, whose axis of revolution is aligned with the $z$ direction and in which the two-sphere swimmer shown in \cref{fig:fig-1-schematic}b is placed at the center. To resolve the stress boundary layers present near the swimmer, we use an unstructured tetrahedral mesh with increasing resolution towards the center of the domain. To ensure results are independent of the choice of mesh size, we have performed mesh convergence studies for each of the tail shapes shown in \cref{fig:different-tail-shapes}. The swimming speed $U$ and rotation rate of the large sphere $\Omega_L$ are determined by advancing \cref{nd_stokes} forward in time, solving for $U_z$ and $\Omega_L$ at each time step such that $F_z=0$ and $L_z=0$ via a quasi-Newton iteration - specifically Broyden's method \citep{broyden1965class}. 

\subsection{Secondary effects on the swimming speed: confinement and shear-thinning}
\label{sec:conf-and-st}
\begin{figure}
    \centering
    \includegraphics[width=\linewidth]{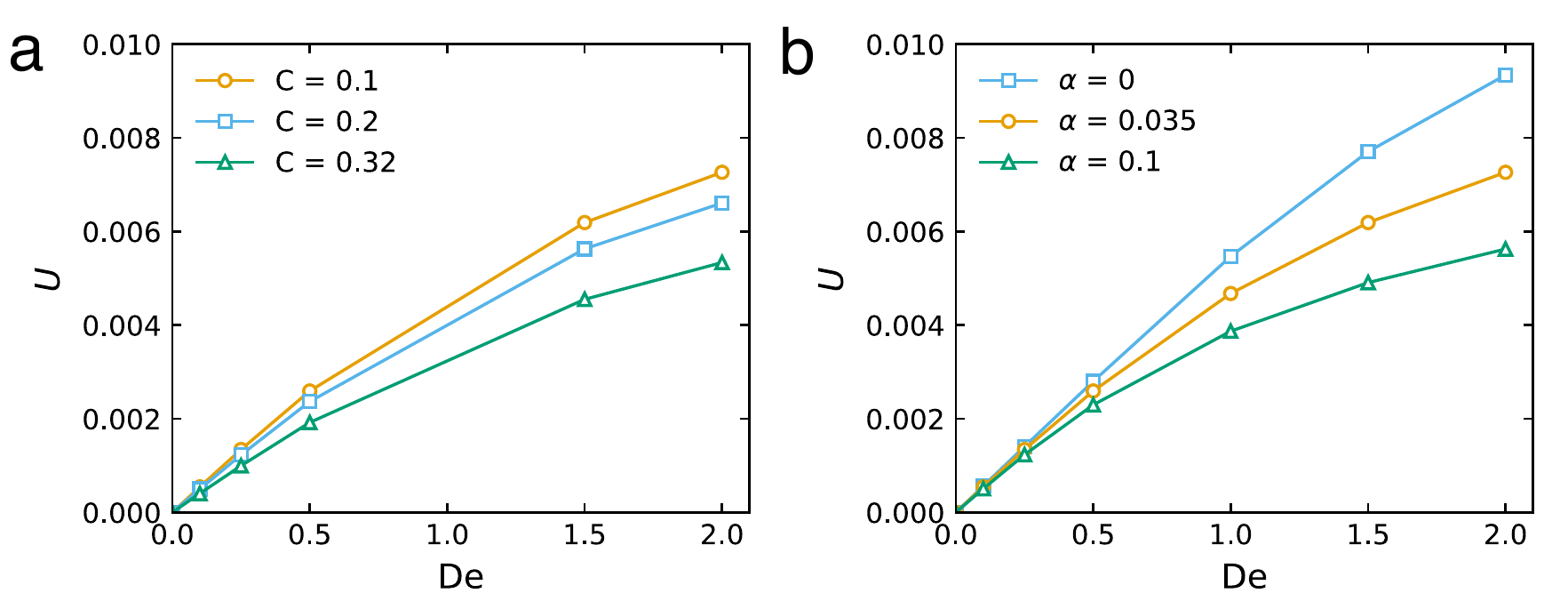}
    \caption{(a) Effect of confinement $C=2R_L/W$ on the swimming speed $U$. Simulations were performed using the single-mode Giesekus constitutive equation with $\alpha_m=0.035$, $\beta=0.46$, $\rst=0.43$, and $\hst=0.17$. As the confinement around the swimmer increases (i.e. increasing $C$), the swimming speed is observed to decrease. (b) Effect of the Giesekus mobility parameter $\alpha_m$ on the swimming speed $U$. For nonzero values of $\alpha_m$, the Giesekus equation exhibits shear-thinning and a nonzero second normal stress coefficient. Simulations for this subfigure were performed using the single-mode Giesekus constitutive equation with $C=0.1$, $\beta=0.46$, $\rst=0.43$, and $\hst=0.17$. We find that as $\alpha_m$ increases, the speed of the swimmer decreases for all De.}
    \label{fig:conf-and-st}
\end{figure}

As discussed in the main text, two effects not considered in the asymptotic theory that were found to be necessary in the numerical simulations for favorable agreement with the experimental data were realistic shear-thinning rheology and confinement of the swimmer. From \cref{fig:conf-and-st}a, we that as the confinement around the swimmer $C=2R_L/W$ increases, the swimming speed decreases for all De. From \cref{fig:conf-and-st}b, we see that as $\alpha_m$ increases (corresponding to a larger extent of shear-thinning and larger value of $\Psi_2$ for the fluid), the swimming speed decreases for all De. Note that for $\alpha_m=0$, corresponding to the Oldroyd-B constitutive equation, the fluid exhibits zero shear-thinning and $\Psi_2=0$. In summary, both increasing confinement and increasing the extent of shear-thinning present in the fluid are seen to decrease the swimmer's speed in a viscoelastic fluid.  

\subsection{Formulation and Rheology of the Non-Newtonian Boger Fluid}
\label{sec:rheology}
\begin{figure}[ht!]
    \centering
    \includegraphics[width = \textwidth]{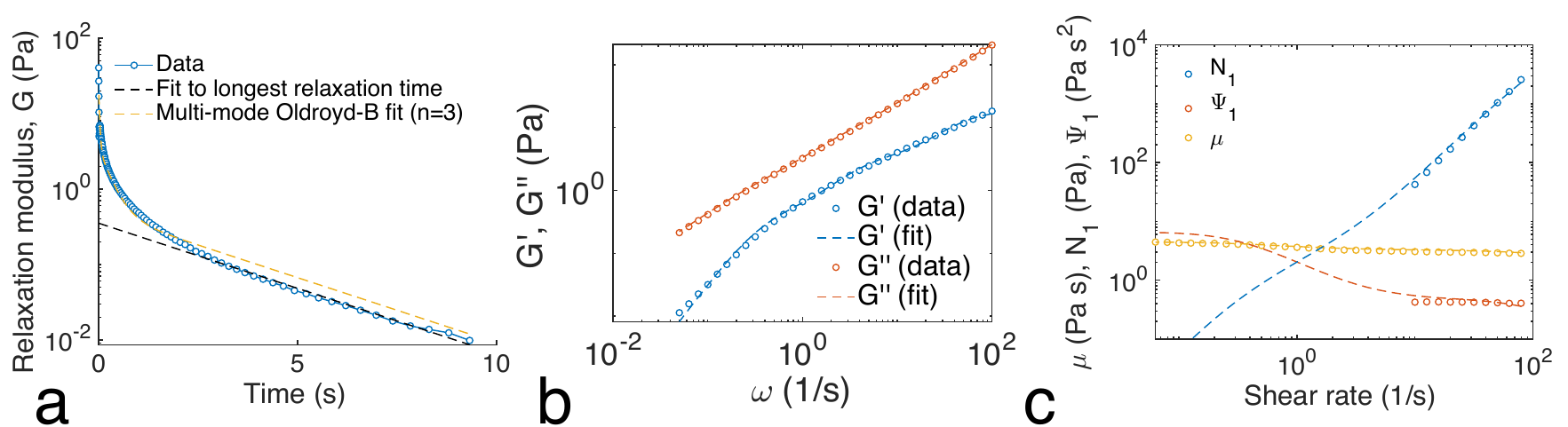}
    \caption{Rheological characterization of the viscoelastic Boger fluid used in this study. (a) Stress relaxation after a step strain of 500\%. Shown alongside the data (blue markers) are a fit to the longest relaxation time of the fluid (black dashed line) and the prediction from the multi-mode Oldroyd-B fit with $n=3$ (yellow dashed line). (b) Small-amplitude oscillatory shear (SAOS) experiment to characterize the linear viscoelastic response of the fluid (i.e. the storage and loss moduli $G'$ and $G''$. Raw data is shown as markers with the multi-mode fit shown as dashed lines. (c) Nonlinear shear rheology of the fluid, i.e. the shear viscosity $\mu$, the first normal stress difference $N_1$ and the first normal stress coefficient $\Psi_1$. The largest shear rate  in the experiments was $\dot\gamma=2\Omega_S\approx 2$ rad/s (i.e. the shear rate present at the equator of the small sphere). The dashed lines denote predictions from the multi-mode Giesekus equation with $\alpha_m^{(0)}=0.2$, $\alpha_m^{(1)}=0.0002$, and $\alpha_m^{(2)}=0.1996$ for the nonlinear rheology (the $\alpha_m^{(i)}$ being determined from a fit to the data for $\Psi_1$).}
    \label{fig:rheology}
\end{figure}

A model viscoelastic fluid, i.e. a "Boger" fluid \citep{boger2012rheological}, was prepared to minimize shear-thinning in the rheological properties and thus to isolate the effects of fluid elasticity on propulsion. The composition of our Boger fluid was 92.5\% corn syrup (Karo syrup), 7.4\% deionized water, and 0.1\% polyacrylamide (MW = $5\times 10^6$ g/mol) by weight. The rheology of the fluid was measured using an ARES-G2 rheometer and is shown in \cref{fig:rheology}. Several experiments were conducted to characterize the fluid on an ARES-G2 rheometer, including stress relaxation after a step strain (\cref{fig:rheology}a), small-amplitude oscillatory shear (\cref{fig:rheology}b), and steady shear (\cref{fig:rheology}c). 
Stress relaxation after a step strain (\cref{fig:rheology}a) was used to determine the longest relaxation time of the fluid. This longest relaxation time was found to be $\lambda_{long}\approx 2.51$ s. With the relaxation time determined, a multi-mode Oldroyd-B model was fit to the small-amplitude oscillatory shear (SAOS) data shown in \cref{fig:rheology}b, with the total zero-shear solution viscosity determined by the value of the shear viscosity $\mu$ at the lowest measured shear rate (c.f. \cref{fig:rheology}c). A summary of the fit is provided in \cref{tab:my-table}. This fit yields a solvent viscosity $\mu_s=2.05$ Pa s, a zero-shear solution viscosity $\mu_0=4.46$ Pa s, a viscosity ratio $\beta=0.46$, and an average relaxation time $\bar{\lambda}=1.38$ s. In situations where we wish to model shear-thinning in the fluid and a nonzero second normal stress coefficient $\Psi_2$, we fit the shear rheology shown in \cref{fig:rheology}c to either a single-mode or multi-mode Giesekus model. The fit to the single-mode Giesekus equation (which was fit to the shear viscosity) yields a best fit value of $\alpha_m=0.035$ and the fit to the multi-mode Giesekus equation with $n=3$ (which was fit to the data for $\Psi_1$) yields $\alpha_m^{(0)}=0.2$, $\alpha_m^{(1)}=0.0002$, and $\alpha_m^{(2)}=0.1996$. This fit is shown as the yellow dashed line in \cref{fig:rheology}c. Note that for the single-mode Giesekus equation, $B=-2\Psi_2/\Psi_1=\alpha_m$ \citep{bird1987dynamics}, showing that a fluid with a nonzero value of $\alpha_m$ not only exhibits shear-thinning but also exhibits nonzero second normal stress difference. Lastly, a power-law fluid model was fit to the shear viscosity data presented in \cref{fig:rheology}c to ascertain the degree of shear-thinning present in the fluid. A shear-thinning exponent of 0.93 was found, indicating a very modest degree of shear-thinning present in the fluid. Based on this and the fact that the the largest shear rate experienced in the robot experiments is in fact $\dot\gamma=2\Omega_S\approx 2$ rad/s (i.e. the shear rate present at the equator of the small sphere), we concluded that this fluid meets our criteria for an elastic fluid with minimal shear-thinning.

\begin{table}[]
\begin{tabular}{|c|c|c|}
\hline
Mode & Viscosity (Pa s) & Relaxation time (s) \\ \hline
0    & 1.23             & 2.51                \\ \hline
1    & 0.82             & 0.28                \\ \hline
2    & 0.37             & 0.03                \\ \hline
\end{tabular}
\caption{Summary of the multi-mode fit to the rheological data presented in \cref{fig:rheology}. Note that these values of $\lambda^{(i)}$ and $\mu^{(i)}_p$ are relevant for both a multi-mode Oldroyd-B and Giesekus fit since the two predict the same linear viscoelastic response (i.e. $G'$ and $G''$). The multi-mode Giesekus fit consits of these parameters plus the following values for the mobility parameters for each mode (determined from a fit to the nonlinear rheology, \cref{fig:rheology}c): $\alpha_m^{(0)}=0.2$, $\alpha_m^{(1)}=0.0002$, and $\alpha_m^{(2)}=0.1996$. }
\label{tab:my-table}
\end{table}

\subsection{Tail Shape Optimization Procedure}
\label{sec:tail-optimization}
To complete optimization of propulsion for tail shapes of various geometries, a mesh and timestep convergence study was used to determine the coarsest spatial and temporal dimension that could give a translation speed $U$ within 15\% of the finest grid result. To run the geometric optimization, many function evaluations are needed, and so running many simulations at the finest mesh resolution is infeasible. 
To generate an initial guess for the optimizer, univariate optimizations were run for all the geometric degrees of freedom (i.e., radius and height of a cylinder). A Sequential Least-Squares Programming (SLSP) method was then used with the initial guess to find the global maximum, with a convergence tolerance of $1\times 10^{-4}$ with respect to $U$. The SLSP method estimates the Jacobian of the objective function at a given point to converge quickly to the function's minimum. To confirm the coarse grid results, multiple points were chosen nearby the found maximum and were run both on the coarse and fine mesh. This data confirmed that the found maximum on the coarse grid was also a maximum on the fine grid within the chosen convergence tolerance. After the optimization, a sweep of different geometric values was chosen to evaluate the global characteristics of the objective function $U$ and to confirm that the global maximum was found (see \cref{sec:tail-optimization}).

\subsection{Maintaining Neutral Buoyancy}
\begin{figure}[h]
\centering
\includegraphics[width=0.85\textwidth]{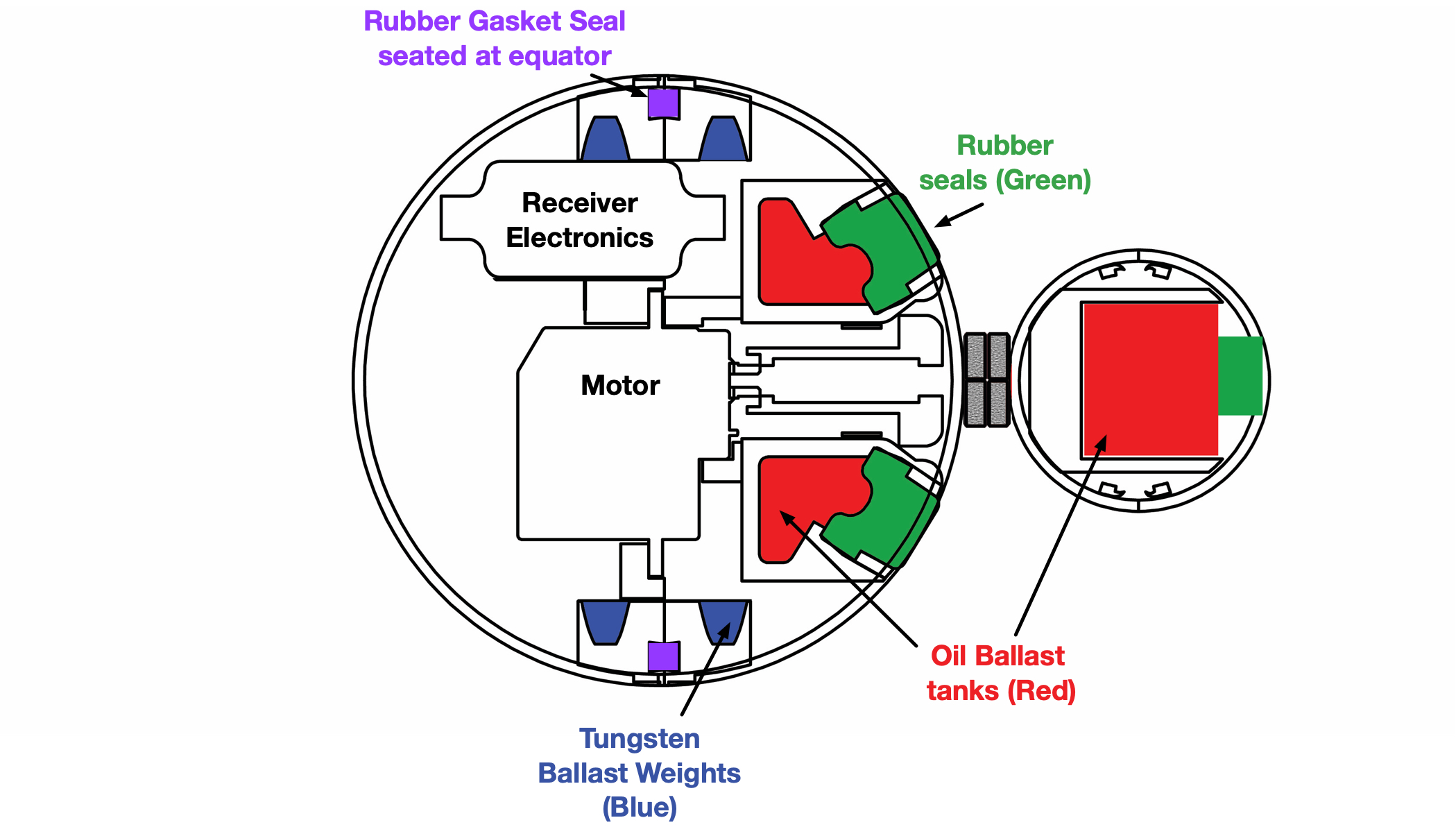}
\caption{\label{fig:ballast} Methods for bouyancy tuning are pictured. Oil ballast tanks (red) are used for fine tuning during experiments, by the user injecting small volumes of oil into the tanks (displacing air) through a rubber stopper. Modular tungsten weights are used in the head for large adjustments, but these large adjustments require a moderate amount of disassembly. }
\end{figure}

One important design aspect of this device is a practical way to easily/precisely tune its mass. 
Because the forward motion of the swimmer is a crucial measurement signal, it is necessary to either know the sum of the forces acting on the robot due to gravity and buoyancy \textit{a priori} -- or if this information is not known, it is necessary to tune these forces to be as close to zero as possible prior to use. 

In our prototype device, we tune the mass of the robot to achieve neutral buoyancy in two stages: gross and fine. In the first stage of gross tuning, the head of the robot is taken apart into two hemispheres. The sphere comes apart in these 2 discrete halves, with a (compressible) rubber gasket at the equator. In each half, there are slots along the interior for tungsten ballast weights, and tungsten putty. This allows tuning of mass in the range of approximately of 20-60g from the initial base weight. This gross mass change allows the device to be tuned for a variety of submersion fluids that vary widely in density.

In the second stage of buoyancy tuning, fine adjustments can be made (on the order of ~0.1 grams). This is achieved by adding oil to small tanks on the device. After gross-adjustments (when the device is fully assembled and often submerged in the fluid), a sharp needle is used to penetrate a rubber "vacutainer" seal, and oil can be added in the tail or to two locations in the head. The seal self-heals when the sharp is removed and can be used for thousands of use cycles. 

As the oil tanks themselves are non-compliant, very small and water-tight, a release needle (to allow for air to escape) is also necessary when adding fluid to the oil ballasts.  

This two-stage method allows for very fine control over the buoyancy of the laboratory prototype, which is necessary to achieve good measurement signal. The same 2-step buoyancy tuning method is also used on the tail attachments. 

It is proposed that active buoyancy control would be ideal in this circumstance, although difficult to achieve on such small-scale prototypes (automatically self-adjusting the mass of the swimmer while in use, based on the density of the submersion fluid). 

For very small-scale device versions, it seems feasible to manufacture a series of devices with unique densities -- and subsequently choose the most appropriate device based on the submersion fluid. 

\subsection{Center of Mass Tuning: Trim and Stability}
\begin{figure}[h]
\centering
\includegraphics[width=0.85\textwidth]{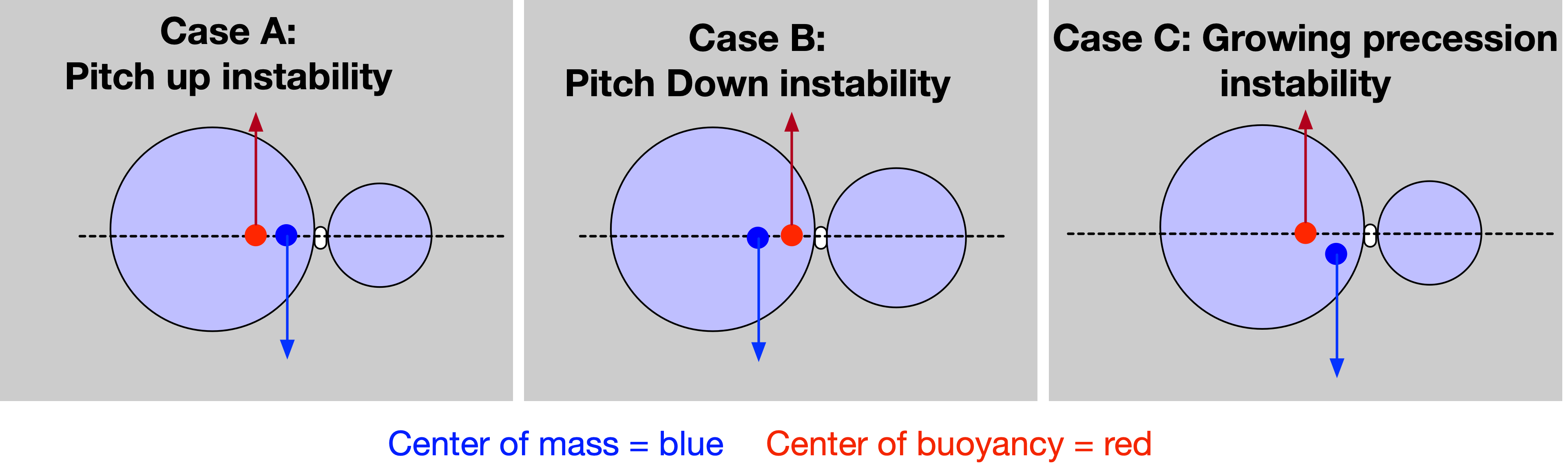}
\caption{\label{fig:stability} Modes of swimmer instability are pictured above.}
\end{figure}
In addition to tuning the overall mass of the swimmer, it is necessary to consider the the exact position of the center of mass in the robot. 

Another way of phrasing this: it is ideal if the distance (along the swimming axis) between the center of gravity and the center of buoyancy is minimized. The center of gravity will be located at the center of mass of the robot (including the mass of the magnetic linker, which is not negligible). The center of buoyancy will be located at the geometric/volumetric center of the robot in a homogeneous submersion fluid; and more generally, at the center of mass of the fluid the device is displacing. 

Even if the \textit{total} mass of the robot is nearly perfectly tuned, the tail will have a tendency to pitch up or down relative to the forward motion vector of the device (especially once the device begins to propel forward), if the \textit{relative} mass between the head and tail has not also been considered. The presence of these instabilities (further detailed in \cref{fig:stability}) decreases/obfuscates the apparent forward propulsion, lowering the accuracy of the device -- if not carefully taken into account.

Investigations into the stability of the swimmer are ongoing. We have observed experimentally that tails of larger diameters (0.5 time the head sphere and larger) are far more sensitive to this effect than smaller diameter tails.  

\begin{figure}[ht!]
    \centering
    \includegraphics[width = 0.8\textwidth]{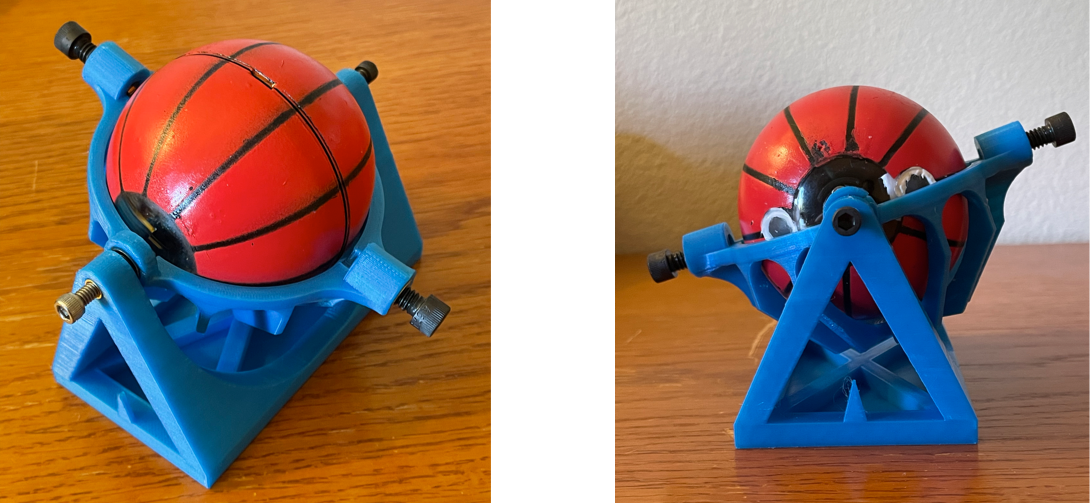}
    \caption{Balancing rig is pictured for aligning the head's center of mass along the swimming axis}
    \label{fig:balance}
\end{figure}

The balancing rig, pictured in \cref{fig:balance}, was used just prior to experimental data collection to ensure that the center of mass within the head was precisely centered to avoid case the unstable case (c) in \cref{fig:stability}. 

\label{sec:balance}

This sensitivity to center of mass (on and off-axis) is challenging to overcome in horizontal swimming modes. We have now begun to move toward vertical swimming modalities where the swimmer swims downward in the fluid, such that the center of mass and center of buoyancy are easier to tune on-axis (this data is not pictured). 
\newpage
\subsection{Schlieren Imaging of Fluid Flow}
\label{sec:schlieren}
\begin{figure}[ht!]
    \centering
    \includegraphics[width = 0.8\textwidth]{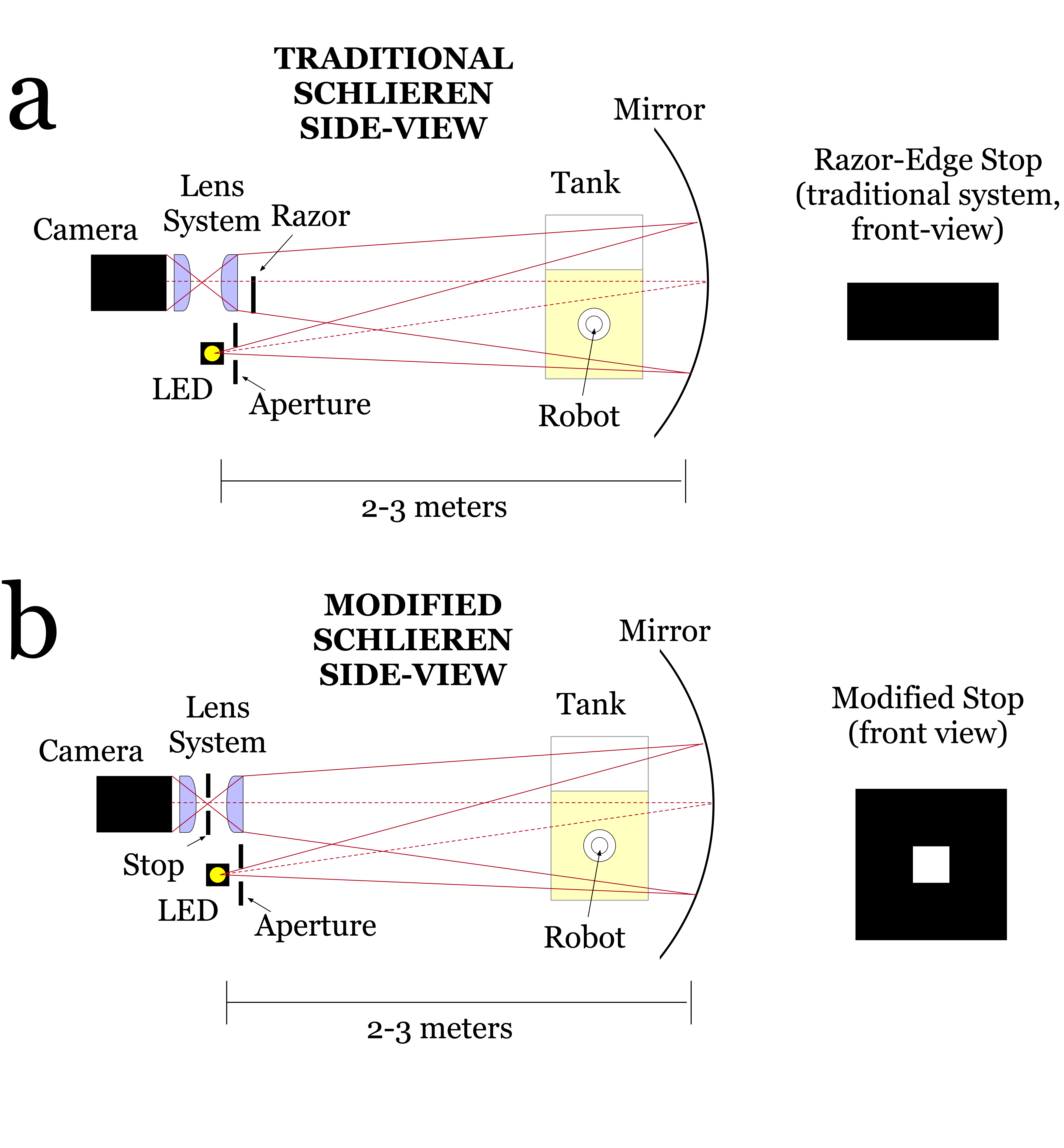}
    \caption{Imaging system for capturing fluid features in the Boger Fluid}
    \label{fig:schlieren}
\end{figure}

A slightly modified Schlieren imaging system was constructed to visualize features in the flow, shown in \cref{fig:jetformation}. 

The use of multiple edges at the stop (instead of a single razor blade) allows us to see gradients along both x and y axes. 

The refracted angle of a beam interacting with a finite element at x,y can then be written as the sum:

\begin{equation}
    \theta = \dfrac{1}{n} \int ( | \dfrac{\partial n_{x,y}}{\partial y}
  |  +| \dfrac{\partial n_{x,y}}{\partial x} |  ) dz
\end{equation}

Where $n_{x,y}$ is the index of refraction in a specific finite element of the tank. The contrast in \cref{fig:jetformation} is a threshold function of the refracted angle, $\theta$. Edges on both sides of the aperture allow both positive and negative gradients along that axis to achieve the same effect in refraction angle (absolute values on terms). 

Changes in the index of refraction in this system may be caused by several factors. We expect that minor heterogeneity in the fluid flow (which will tend to propagate along the flow direction) will visually highlight the flow structures. This is a well-known property of water-based Boger fluids, however has a very minimal rheological effect (for example, samples from different parts of the tank will have indistinguishably similar storage and loss moduli). Another possibility is that the polymer stress is inducing meso-scale lensing in the fluid, leading to regions that tend to appear darker where large changes in polymer stress are present (hoop stress region, viscoelastic jet). The latter possibility is under further study within our group, and warrants further careful investigation. 

\section*{Author Contributions}
LAK: Designed and built the robot prototypes (electrical and mechanical); wrote control and data analysis code; collected and analyzed propulsion data; built imaging systems; collected and analyzed Schlieren data; wrote the paper. \\
JPB: characterized and mixed the Boger fluids ; designed and performed numerical simulations; developed asymptotic theory; collected and analyzed propulsion data; wrote the paper.\\
NJE: Wrote data analysis code, ran motion tracking analysis, completed tail optimization analysis, collected and analyzed propulsion and Schlieren data\\
MP: Initiated original concept; contributed to design of robot; guided the investigation; wrote and edited the paper. \\
ESGS: Initiated original concept; contributed to design and guided the investigation; developed asymptotic theory; wrote and edited the paper. 

\section*{Acknowledgements}
M.P. thanks the Gordon and Betty Moore Foundation (grant no. 5762) and the NSF/UCSF award for the Center for Cellular Construction project (grant no. 9917sc). E.S.G.S. acknowledges that this work is also supported in part by the National Science Foundation (NSF)  grant no. CBET 1803765. J.P.B. acknowledgements support from the Stanford Office of the Vice Provost for Graduation Education and the Gerald J. Lieberman fellowship. LAK and NE thank Dr. G.K. Herring for assistance with Schlieren imaging alignment. The authors also thank Prof. Gerald Fuller for his helpful informal discussions on polymer flow imaging.

\section*{Competing Interests}
The authors currently have no competing financial interests. A patent application \citep{shaqfeh_prakash_kroo_binagia} has been recently filed based on this work, (Assignee: The Board of Trustees of the Leland Stanford Junior University).

\bibliography{swimmerbib} 

\begin{thebibliography}{39}%
\makeatletter
\providecommand \@ifxundefined [1]{%
 \@ifx{#1\undefined}
}%
\providecommand \@ifnum [1]{%
 \ifnum #1\expandafter \@firstoftwo
 \else \expandafter \@secondoftwo
 \fi
}%
\providecommand \@ifx [1]{%
 \ifx #1\expandafter \@firstoftwo
 \else \expandafter \@secondoftwo
 \fi
}%
\providecommand \natexlab [1]{#1}%
\providecommand \enquote  [1]{``#1''}%
\providecommand \bibnamefont  [1]{#1}%
\providecommand \bibfnamefont [1]{#1}%
\providecommand \citenamefont [1]{#1}%
\providecommand \href@noop [0]{\@secondoftwo}%
\providecommand \href [0]{\begingroup \@sanitize@url \@href}%
\providecommand \@href[1]{\@@startlink{#1}\@@href}%
\providecommand \@@href[1]{\endgroup#1\@@endlink}%
\providecommand \@sanitize@url [0]{\catcode `\\12\catcode `\$12\catcode
  `\&12\catcode `\#12\catcode `\^12\catcode `\_12\catcode `\%12\relax}%
\providecommand \@@startlink[1]{}%
\providecommand \@@endlink[0]{}%
\providecommand \url  [0]{\begingroup\@sanitize@url \@url }%
\providecommand \@url [1]{\endgroup\@href {#1}{\urlprefix }}%
\providecommand \urlprefix  [0]{URL }%
\providecommand \Eprint [0]{\href }%
\providecommand \doibase [0]{https://doi.org/}%
\providecommand \selectlanguage [0]{\@gobble}%
\providecommand \bibinfo  [0]{\@secondoftwo}%
\providecommand \bibfield  [0]{\@secondoftwo}%
\providecommand \translation [1]{[#1]}%
\providecommand \BibitemOpen [0]{}%
\providecommand \bibitemStop [0]{}%
\providecommand \bibitemNoStop [0]{.\EOS\space}%
\providecommand \EOS [0]{\spacefactor3000\relax}%
\providecommand \BibitemShut  [1]{\csname bibitem#1\endcsname}%
\let\auto@bib@innerbib\@empty
\bibitem [{\citenamefont {Webb}\ \emph {et~al.}(2001)\citenamefont {Webb},
  \citenamefont {Simonetti},\ and\ \citenamefont {Jones}}]{webb2001slocum}%
  \BibitemOpen
  \bibfield  {author} {\bibinfo {author} {\bibfnamefont {D.~C.}\ \bibnamefont
  {Webb}}, \bibinfo {author} {\bibfnamefont {P.~J.}\ \bibnamefont
  {Simonetti}},\ and\ \bibinfo {author} {\bibfnamefont {C.~P.}\ \bibnamefont
  {Jones}},\ }\bibfield  {title} {\bibinfo {title} {Slocum: An underwater
  glider propelled by environmental energy},\ }\href@noop {} {\bibfield
  {journal} {\bibinfo  {journal} {IEEE Journal of oceanic engineering}\
  }\textbf {\bibinfo {volume} {26}},\ \bibinfo {pages} {447} (\bibinfo {year}
  {2001})}\BibitemShut {NoStop}%
\bibitem [{\citenamefont {Parness}\ \emph {et~al.}(2009)\citenamefont
  {Parness}, \citenamefont {Soto}, \citenamefont {Esparza}, \citenamefont
  {Gravish}, \citenamefont {Wilkinson}, \citenamefont {Autumn},\ and\
  \citenamefont {Cutkosky}}]{parness2009microfabricated}%
  \BibitemOpen
  \bibfield  {author} {\bibinfo {author} {\bibfnamefont {A.}~\bibnamefont
  {Parness}}, \bibinfo {author} {\bibfnamefont {D.}~\bibnamefont {Soto}},
  \bibinfo {author} {\bibfnamefont {N.}~\bibnamefont {Esparza}}, \bibinfo
  {author} {\bibfnamefont {N.}~\bibnamefont {Gravish}}, \bibinfo {author}
  {\bibfnamefont {M.}~\bibnamefont {Wilkinson}}, \bibinfo {author}
  {\bibfnamefont {K.}~\bibnamefont {Autumn}},\ and\ \bibinfo {author}
  {\bibfnamefont {M.}~\bibnamefont {Cutkosky}},\ }\bibfield  {title} {\bibinfo
  {title} {A microfabricated wedge-shaped adhesive array displaying gecko-like
  dynamic adhesion, directionality and long lifetime},\ }\href@noop {}
  {\bibfield  {journal} {\bibinfo  {journal} {Journal of the Royal Society
  Interface}\ }\textbf {\bibinfo {volume} {6}},\ \bibinfo {pages} {1223}
  (\bibinfo {year} {2009})}\BibitemShut {NoStop}%
\bibitem [{\citenamefont {Spagnolie}(2015)}]{spagnolie2015complex}%
  \BibitemOpen
  \bibfield  {author} {\bibinfo {author} {\bibfnamefont {S.~E.}\ \bibnamefont
  {Spagnolie}},\ }\bibfield  {title} {\bibinfo {title} {Complex fluids in
  biological systems},\ }\href@noop {} {\bibfield  {journal} {\bibinfo
  {journal} {Biological and Medical Physics, Biomedical Engineering}\ }
  (\bibinfo {year} {2015})}\BibitemShut {NoStop}%
\bibitem [{\citenamefont {Gilpin}\ \emph {et~al.}(2017)\citenamefont {Gilpin},
  \citenamefont {Prakash},\ and\ \citenamefont {Prakash}}]{gilpin2017vortex}%
  \BibitemOpen
  \bibfield  {author} {\bibinfo {author} {\bibfnamefont {W.}~\bibnamefont
  {Gilpin}}, \bibinfo {author} {\bibfnamefont {V.~N.}\ \bibnamefont
  {Prakash}},\ and\ \bibinfo {author} {\bibfnamefont {M.}~\bibnamefont
  {Prakash}},\ }\bibfield  {title} {\bibinfo {title} {Vortex arrays and ciliary
  tangles underlie the feeding--swimming trade-off in starfish larvae},\
  }\href@noop {} {\bibfield  {journal} {\bibinfo  {journal} {Nature Physics}\
  }\textbf {\bibinfo {volume} {13}},\ \bibinfo {pages} {380} (\bibinfo {year}
  {2017})}\BibitemShut {NoStop}%
\bibitem [{\citenamefont {Bull}\ \emph {et~al.}(2021)\citenamefont {Bull},
  \citenamefont {Kroo},\ and\ \citenamefont {Prakash}}]{bull2021excitable}%
  \BibitemOpen
  \bibfield  {author} {\bibinfo {author} {\bibfnamefont {M.~S.}\ \bibnamefont
  {Bull}}, \bibinfo {author} {\bibfnamefont {L.~A.}\ \bibnamefont {Kroo}},\
  and\ \bibinfo {author} {\bibfnamefont {M.}~\bibnamefont {Prakash}},\
  }\bibfield  {title} {\bibinfo {title} {Excitable mechanics embodied in a
  walking cilium},\ }\href@noop {} {\bibfield  {journal} {\bibinfo  {journal}
  {arXiv preprint arXiv:2107.02930}\ } (\bibinfo {year} {2021})}\BibitemShut
  {NoStop}%
\bibitem [{\citenamefont {Celli}\ \emph {et~al.}(2009)\citenamefont {Celli},
  \citenamefont {Turner}, \citenamefont {Afdhal}, \citenamefont {Keates},
  \citenamefont {Ghiran}, \citenamefont {Kelly}, \citenamefont {Ewoldt},
  \citenamefont {McKinley}, \citenamefont {So}, \citenamefont {Erramilli} \emph
  {et~al.}}]{celli2009helicobacter}%
  \BibitemOpen
  \bibfield  {author} {\bibinfo {author} {\bibfnamefont {J.~P.}\ \bibnamefont
  {Celli}}, \bibinfo {author} {\bibfnamefont {B.~S.}\ \bibnamefont {Turner}},
  \bibinfo {author} {\bibfnamefont {N.~H.}\ \bibnamefont {Afdhal}}, \bibinfo
  {author} {\bibfnamefont {S.}~\bibnamefont {Keates}}, \bibinfo {author}
  {\bibfnamefont {I.}~\bibnamefont {Ghiran}}, \bibinfo {author} {\bibfnamefont
  {C.~P.}\ \bibnamefont {Kelly}}, \bibinfo {author} {\bibfnamefont {R.~H.}\
  \bibnamefont {Ewoldt}}, \bibinfo {author} {\bibfnamefont {G.~H.}\
  \bibnamefont {McKinley}}, \bibinfo {author} {\bibfnamefont {P.}~\bibnamefont
  {So}}, \bibinfo {author} {\bibfnamefont {S.}~\bibnamefont {Erramilli}}, \emph
  {et~al.},\ }\bibfield  {title} {\bibinfo {title} {Helicobacter pylori moves
  through mucus by reducing mucin viscoelasticity},\ }\href@noop {} {\bibfield
  {journal} {\bibinfo  {journal} {Proceedings of the National Academy of
  Sciences}\ }\textbf {\bibinfo {volume} {106}},\ \bibinfo {pages} {14321}
  (\bibinfo {year} {2009})}\BibitemShut {NoStop}%
\bibitem [{\citenamefont {Purcell}(1977)}]{purcell1977life}%
  \BibitemOpen
  \bibfield  {author} {\bibinfo {author} {\bibfnamefont {E.~M.}\ \bibnamefont
  {Purcell}},\ }\bibfield  {title} {\bibinfo {title} {Life at low reynolds
  number},\ }\href@noop {} {\bibfield  {journal} {\bibinfo  {journal} {American
  journal of physics}\ }\textbf {\bibinfo {volume} {45}},\ \bibinfo {pages} {3}
  (\bibinfo {year} {1977})}\BibitemShut {NoStop}%
\bibitem [{\citenamefont {Kim}\ and\ \citenamefont
  {Karrila}(2013)}]{kim2013microhydrodynamics}%
  \BibitemOpen
  \bibfield  {author} {\bibinfo {author} {\bibfnamefont {S.}~\bibnamefont
  {Kim}}\ and\ \bibinfo {author} {\bibfnamefont {S.~J.}\ \bibnamefont
  {Karrila}},\ }\href@noop {} {\emph {\bibinfo {title} {Microhydrodynamics:
  principles and selected applications}}}\ (\bibinfo  {publisher} {Courier
  Corporation},\ \bibinfo {year} {2013})\BibitemShut {NoStop}%
\bibitem [{\citenamefont {Lauga}\ and\ \citenamefont
  {Powers}(2009)}]{lauga2009hydrodynamics}%
  \BibitemOpen
  \bibfield  {author} {\bibinfo {author} {\bibfnamefont {E.}~\bibnamefont
  {Lauga}}\ and\ \bibinfo {author} {\bibfnamefont {T.~R.}\ \bibnamefont
  {Powers}},\ }\bibfield  {title} {\bibinfo {title} {The hydrodynamics of
  swimming microorganisms},\ }\href@noop {} {\bibfield  {journal} {\bibinfo
  {journal} {Reports on Progress in Physics}\ }\textbf {\bibinfo {volume}
  {72}},\ \bibinfo {pages} {096601} (\bibinfo {year} {2009})}\BibitemShut
  {NoStop}%
\bibitem [{\citenamefont {Qiu}\ \emph {et~al.}(2014)\citenamefont {Qiu},
  \citenamefont {Lee}, \citenamefont {Mark}, \citenamefont {Morozov},
  \citenamefont {M{\"u}nster}, \citenamefont {Mierka}, \citenamefont {Turek},
  \citenamefont {Leshansky},\ and\ \citenamefont {Fischer}}]{qiu2014swimming}%
  \BibitemOpen
  \bibfield  {author} {\bibinfo {author} {\bibfnamefont {T.}~\bibnamefont
  {Qiu}}, \bibinfo {author} {\bibfnamefont {T.-C.}\ \bibnamefont {Lee}},
  \bibinfo {author} {\bibfnamefont {A.~G.}\ \bibnamefont {Mark}}, \bibinfo
  {author} {\bibfnamefont {K.~I.}\ \bibnamefont {Morozov}}, \bibinfo {author}
  {\bibfnamefont {R.}~\bibnamefont {M{\"u}nster}}, \bibinfo {author}
  {\bibfnamefont {O.}~\bibnamefont {Mierka}}, \bibinfo {author} {\bibfnamefont
  {S.}~\bibnamefont {Turek}}, \bibinfo {author} {\bibfnamefont {A.~M.}\
  \bibnamefont {Leshansky}},\ and\ \bibinfo {author} {\bibfnamefont
  {P.}~\bibnamefont {Fischer}},\ }\bibfield  {title} {\bibinfo {title}
  {Swimming by reciprocal motion at low reynolds number},\ }\href@noop {}
  {\bibfield  {journal} {\bibinfo  {journal} {Nature communications}\ }\textbf
  {\bibinfo {volume} {5}},\ \bibinfo {pages} {1} (\bibinfo {year}
  {2014})}\BibitemShut {NoStop}%
\bibitem [{\citenamefont {Keim}\ \emph {et~al.}(2012)\citenamefont {Keim},
  \citenamefont {Garcia},\ and\ \citenamefont {Arratia}}]{keim2012fluid}%
  \BibitemOpen
  \bibfield  {author} {\bibinfo {author} {\bibfnamefont {N.~C.}\ \bibnamefont
  {Keim}}, \bibinfo {author} {\bibfnamefont {M.}~\bibnamefont {Garcia}},\ and\
  \bibinfo {author} {\bibfnamefont {P.~E.}\ \bibnamefont {Arratia}},\
  }\bibfield  {title} {\bibinfo {title} {Fluid elasticity can enable propulsion
  at low reynolds number},\ }\href@noop {} {\bibfield  {journal} {\bibinfo
  {journal} {Physics of Fluids}\ }\textbf {\bibinfo {volume} {24}},\ \bibinfo
  {pages} {081703} (\bibinfo {year} {2012})}\BibitemShut {NoStop}%
\bibitem [{\citenamefont {Datt}\ \emph {et~al.}(2018)\citenamefont {Datt},
  \citenamefont {Nasouri},\ and\ \citenamefont {Elfring}}]{datt2018two}%
  \BibitemOpen
  \bibfield  {author} {\bibinfo {author} {\bibfnamefont {C.}~\bibnamefont
  {Datt}}, \bibinfo {author} {\bibfnamefont {B.}~\bibnamefont {Nasouri}},\ and\
  \bibinfo {author} {\bibfnamefont {G.~J.}\ \bibnamefont {Elfring}},\
  }\bibfield  {title} {\bibinfo {title} {Two-sphere swimmers in viscoelastic
  fluids},\ }\href@noop {} {\bibfield  {journal} {\bibinfo  {journal} {Physical
  Review Fluids}\ }\textbf {\bibinfo {volume} {3}},\ \bibinfo {pages} {123301}
  (\bibinfo {year} {2018})}\BibitemShut {NoStop}%
\bibitem [{\citenamefont {Pak}\ \emph {et~al.}(2012)\citenamefont {Pak},
  \citenamefont {Zhu}, \citenamefont {Brandt},\ and\ \citenamefont
  {Lauga}}]{pak2012micropropulsion}%
  \BibitemOpen
  \bibfield  {author} {\bibinfo {author} {\bibfnamefont {O.~S.}\ \bibnamefont
  {Pak}}, \bibinfo {author} {\bibfnamefont {L.}~\bibnamefont {Zhu}}, \bibinfo
  {author} {\bibfnamefont {L.}~\bibnamefont {Brandt}},\ and\ \bibinfo {author}
  {\bibfnamefont {E.}~\bibnamefont {Lauga}},\ }\bibfield  {title} {\bibinfo
  {title} {Micropropulsion and microrheology in complex fluids via symmetry
  breaking},\ }\href@noop {} {\bibfield  {journal} {\bibinfo  {journal}
  {Physics of fluids}\ }\textbf {\bibinfo {volume} {24}},\ \bibinfo {pages}
  {103102} (\bibinfo {year} {2012})}\BibitemShut {NoStop}%
\bibitem [{\citenamefont {Puente-Vel{\'a}zquez}\ \emph
  {et~al.}(2019)\citenamefont {Puente-Vel{\'a}zquez}, \citenamefont
  {God{\'\i}nez}, \citenamefont {Lauga},\ and\ \citenamefont
  {Zenit}}]{puente2019viscoelastic}%
  \BibitemOpen
  \bibfield  {author} {\bibinfo {author} {\bibfnamefont {J.~A.}\ \bibnamefont
  {Puente-Vel{\'a}zquez}}, \bibinfo {author} {\bibfnamefont {F.~A.}\
  \bibnamefont {God{\'\i}nez}}, \bibinfo {author} {\bibfnamefont
  {E.}~\bibnamefont {Lauga}},\ and\ \bibinfo {author} {\bibfnamefont
  {R.}~\bibnamefont {Zenit}},\ }\bibfield  {title} {\bibinfo {title}
  {Viscoelastic propulsion of a rotating dumbbell},\ }\href@noop {} {\bibfield
  {journal} {\bibinfo  {journal} {Microfluidics and Nanofluidics}\ }\textbf
  {\bibinfo {volume} {23}},\ \bibinfo {pages} {1} (\bibinfo {year}
  {2019})}\BibitemShut {NoStop}%
\bibitem [{\citenamefont {Binagia}\ and\ \citenamefont
  {Shaqfeh}(2021)}]{binagia2021self}%
  \BibitemOpen
  \bibfield  {author} {\bibinfo {author} {\bibfnamefont {J.~P.}\ \bibnamefont
  {Binagia}}\ and\ \bibinfo {author} {\bibfnamefont {E.~S.}\ \bibnamefont
  {Shaqfeh}},\ }\bibfield  {title} {\bibinfo {title} {Self-propulsion of a
  freely suspended swimmer by a swirling tail in a viscoelastic fluid},\
  }\href@noop {} {\bibfield  {journal} {\bibinfo  {journal} {Physical Review
  Fluids}\ }\textbf {\bibinfo {volume} {6}},\ \bibinfo {pages} {053301}
  (\bibinfo {year} {2021})}\BibitemShut {NoStop}%
\bibitem [{\citenamefont {Oldroyd}(1950)}]{oldroyd1950formulation}%
  \BibitemOpen
  \bibfield  {author} {\bibinfo {author} {\bibfnamefont {J.~G.}\ \bibnamefont
  {Oldroyd}},\ }\bibfield  {title} {\bibinfo {title} {On the formulation of
  rheological equations of state},\ }\href@noop {} {\bibfield  {journal}
  {\bibinfo  {journal} {Proceedings of the Royal Society of London. Series A.
  Mathematical and Physical Sciences}\ }\textbf {\bibinfo {volume} {200}},\
  \bibinfo {pages} {523} (\bibinfo {year} {1950})}\BibitemShut {NoStop}%
\bibitem [{\citenamefont {Bird}\ \emph {et~al.}(1987)\citenamefont {Bird},
  \citenamefont {Armstrong},\ and\ \citenamefont
  {Hassager}}]{bird1987dynamics}%
  \BibitemOpen
  \bibfield  {author} {\bibinfo {author} {\bibfnamefont {R.~B.}\ \bibnamefont
  {Bird}}, \bibinfo {author} {\bibfnamefont {R.~C.}\ \bibnamefont
  {Armstrong}},\ and\ \bibinfo {author} {\bibfnamefont {O.}~\bibnamefont
  {Hassager}},\ }\href@noop {} {\bibinfo {title} {Dynamics of polymeric
  liquids. vol. 1: Fluid mechanics}} (\bibinfo {year} {1987})\BibitemShut
  {NoStop}%
\bibitem [{\citenamefont {Ham}\ \emph {et~al.}(2006)\citenamefont {Ham},
  \citenamefont {Mattsson},\ and\ \citenamefont {Iaccarino}}]{Ham2006}%
  \BibitemOpen
  \bibfield  {author} {\bibinfo {author} {\bibfnamefont {F.}~\bibnamefont
  {Ham}}, \bibinfo {author} {\bibfnamefont {K.}~\bibnamefont {Mattsson}},\ and\
  \bibinfo {author} {\bibfnamefont {G.}~\bibnamefont {Iaccarino}},\ }\bibfield
  {title} {\bibinfo {title} {{Accurate and stable finite volume operators for
  unstructured flow solvers}},\ }\href@noop {} {\bibfield  {journal} {\bibinfo
  {journal} {Center for Turbulence Research Annual Research Briefs}\ ,\
  \bibinfo {pages} {243}} (\bibinfo {year} {2006})}\BibitemShut {NoStop}%
\bibitem [{\citenamefont {Richter}\ \emph {et~al.}(2010)\citenamefont
  {Richter}, \citenamefont {Iaccarino},\ and\ \citenamefont
  {Shaqfeh}}]{Richter2010}%
  \BibitemOpen
  \bibfield  {author} {\bibinfo {author} {\bibfnamefont {D.}~\bibnamefont
  {Richter}}, \bibinfo {author} {\bibfnamefont {G.}~\bibnamefont {Iaccarino}},\
  and\ \bibinfo {author} {\bibfnamefont {E.~S.}\ \bibnamefont {Shaqfeh}},\
  }\bibfield  {title} {\bibinfo {title} {{Simulations of three-dimensional
  viscoelastic flows past a circular cylinder at moderate Reynolds numbers}},\
  }\href {https://doi.org/10.1017/S0022112009994083} {\bibfield  {journal}
  {\bibinfo  {journal} {Journal of Fluid Mechanics}\ }\textbf {\bibinfo
  {volume} {651}},\ \bibinfo {pages} {415} (\bibinfo {year}
  {2010})}\BibitemShut {NoStop}%
\bibitem [{\citenamefont {Padhy}\ \emph {et~al.}(2013)\citenamefont {Padhy},
  \citenamefont {Shaqfeh}, \citenamefont {Iaccarino}, \citenamefont {Morris},\
  and\ \citenamefont {Tonmukayakul}}]{padhy2013simulations}%
  \BibitemOpen
  \bibfield  {author} {\bibinfo {author} {\bibfnamefont {S.}~\bibnamefont
  {Padhy}}, \bibinfo {author} {\bibfnamefont {E.}~\bibnamefont {Shaqfeh}},
  \bibinfo {author} {\bibfnamefont {G.}~\bibnamefont {Iaccarino}}, \bibinfo
  {author} {\bibfnamefont {J.}~\bibnamefont {Morris}},\ and\ \bibinfo {author}
  {\bibfnamefont {N.}~\bibnamefont {Tonmukayakul}},\ }\bibfield  {title}
  {\bibinfo {title} {Simulations of a sphere sedimenting in a viscoelastic
  fluid with cross shear flow},\ }\href@noop {} {\bibfield  {journal} {\bibinfo
   {journal} {Journal of Non-Newtonian Fluid Mechanics}\ }\textbf {\bibinfo
  {volume} {197}},\ \bibinfo {pages} {48} (\bibinfo {year} {2013})}\BibitemShut
  {NoStop}%
\bibitem [{\citenamefont {Yang}\ \emph {et~al.}(2016)\citenamefont {Yang},
  \citenamefont {Krishnan},\ and\ \citenamefont {Shaqfeh}}]{yang2016numerical}%
  \BibitemOpen
  \bibfield  {author} {\bibinfo {author} {\bibfnamefont {M.}~\bibnamefont
  {Yang}}, \bibinfo {author} {\bibfnamefont {S.}~\bibnamefont {Krishnan}},\
  and\ \bibinfo {author} {\bibfnamefont {E.~S.}\ \bibnamefont {Shaqfeh}},\
  }\bibfield  {title} {\bibinfo {title} {Numerical simulations of the rheology
  of suspensions of rigid spheres at low volume fraction in a viscoelastic
  fluid under shear},\ }\href@noop {} {\bibfield  {journal} {\bibinfo
  {journal} {Journal of Non-Newtonian Fluid Mechanics}\ }\textbf {\bibinfo
  {volume} {233}},\ \bibinfo {pages} {181} (\bibinfo {year}
  {2016})}\BibitemShut {NoStop}%
\bibitem [{\citenamefont {Saadat}\ \emph {et~al.}(2018)\citenamefont {Saadat},
  \citenamefont {Guido}, \citenamefont {Iaccarino},\ and\ \citenamefont
  {Shaqfeh}}]{saadat2018immersed}%
  \BibitemOpen
  \bibfield  {author} {\bibinfo {author} {\bibfnamefont {A.}~\bibnamefont
  {Saadat}}, \bibinfo {author} {\bibfnamefont {C.~J.}\ \bibnamefont {Guido}},
  \bibinfo {author} {\bibfnamefont {G.}~\bibnamefont {Iaccarino}},\ and\
  \bibinfo {author} {\bibfnamefont {E.~S.}\ \bibnamefont {Shaqfeh}},\
  }\bibfield  {title} {\bibinfo {title} {Immersed-finite-element method for
  deformable particle suspensions in viscous and viscoelastic media},\
  }\href@noop {} {\bibfield  {journal} {\bibinfo  {journal} {Physical Review
  E}\ }\textbf {\bibinfo {volume} {98}},\ \bibinfo {pages} {063316} (\bibinfo
  {year} {2018})}\BibitemShut {NoStop}%
\bibitem [{\citenamefont {Binagia}\ \emph {et~al.}(2019)\citenamefont
  {Binagia}, \citenamefont {Guido},\ and\ \citenamefont
  {Shaqfeh}}]{binagia2019three}%
  \BibitemOpen
  \bibfield  {author} {\bibinfo {author} {\bibfnamefont {J.~P.}\ \bibnamefont
  {Binagia}}, \bibinfo {author} {\bibfnamefont {C.~J.}\ \bibnamefont {Guido}},\
  and\ \bibinfo {author} {\bibfnamefont {E.~S.}\ \bibnamefont {Shaqfeh}},\
  }\bibfield  {title} {\bibinfo {title} {Three-dimensional simulations of
  undulatory and amoeboid swimmers in viscoelastic fluids},\ }\href@noop {}
  {\bibfield  {journal} {\bibinfo  {journal} {Soft matter}\ }\textbf {\bibinfo
  {volume} {15}},\ \bibinfo {pages} {4836} (\bibinfo {year}
  {2019})}\BibitemShut {NoStop}%
\bibitem [{\citenamefont {Binagia}\ \emph {et~al.}(2020)\citenamefont
  {Binagia}, \citenamefont {Phoa}, \citenamefont {Housiadas},\ and\
  \citenamefont {Shaqfeh}}]{binagia2020swimming}%
  \BibitemOpen
  \bibfield  {author} {\bibinfo {author} {\bibfnamefont {J.~P.}\ \bibnamefont
  {Binagia}}, \bibinfo {author} {\bibfnamefont {A.}~\bibnamefont {Phoa}},
  \bibinfo {author} {\bibfnamefont {K.~D.}\ \bibnamefont {Housiadas}},\ and\
  \bibinfo {author} {\bibfnamefont {E.~S.}\ \bibnamefont {Shaqfeh}},\
  }\bibfield  {title} {\bibinfo {title} {Swimming with swirl in a viscoelastic
  fluid},\ }\href@noop {} {\bibfield  {journal} {\bibinfo  {journal} {Journal
  of Fluid Mechanics}\ }\textbf {\bibinfo {volume} {900}} (\bibinfo {year}
  {2020})}\BibitemShut {NoStop}%
\bibitem [{\citenamefont {Housiadas}\ \emph {et~al.}(2021)\citenamefont
  {Housiadas}, \citenamefont {Binagia},\ and\ \citenamefont
  {Shaqfeh}}]{housiadas_binagia_shaqfeh_2021}%
  \BibitemOpen
  \bibfield  {author} {\bibinfo {author} {\bibfnamefont {K.~D.}\ \bibnamefont
  {Housiadas}}, \bibinfo {author} {\bibfnamefont {J.~P.}\ \bibnamefont
  {Binagia}},\ and\ \bibinfo {author} {\bibfnamefont {E.~S.}\ \bibnamefont
  {Shaqfeh}},\ }\bibfield  {title} {\bibinfo {title} {Squirmers with swirl at
  low weissenberg number},\ }\href {https://doi.org/10.1017/jfm.2020.987}
  {\bibfield  {journal} {\bibinfo  {journal} {Journal of Fluid Mechanics}\
  }\textbf {\bibinfo {volume} {911}},\ \bibinfo {pages} {A16} (\bibinfo {year}
  {2021})}\BibitemShut {NoStop}%
\bibitem [{\citenamefont {Broyden}(1965)}]{broyden1965class}%
  \BibitemOpen
  \bibfield  {author} {\bibinfo {author} {\bibfnamefont {C.~G.}\ \bibnamefont
  {Broyden}},\ }\bibfield  {title} {\bibinfo {title} {A class of methods for
  solving nonlinear simultaneous equations},\ }\href@noop {} {\bibfield
  {journal} {\bibinfo  {journal} {Mathematics of computation}\ }\textbf
  {\bibinfo {volume} {19}},\ \bibinfo {pages} {577} (\bibinfo {year}
  {1965})}\BibitemShut {NoStop}%
\bibitem [{\citenamefont {Boger}\ and\ \citenamefont
  {Walters}(2012)}]{boger2012rheological}%
  \BibitemOpen
  \bibfield  {author} {\bibinfo {author} {\bibfnamefont {D.~V.}\ \bibnamefont
  {Boger}}\ and\ \bibinfo {author} {\bibfnamefont {K.}~\bibnamefont
  {Walters}},\ }\href@noop {} {\emph {\bibinfo {title} {Rheological phenomena
  in focus}}}\ (\bibinfo  {publisher} {Elsevier},\ \bibinfo {year}
  {2012})\BibitemShut {NoStop}%
\bibitem [{\citenamefont {Morrison}\ \emph {et~al.}(2001)\citenamefont
  {Morrison} \emph {et~al.}}]{morrison2001understanding}%
  \BibitemOpen
  \bibfield  {author} {\bibinfo {author} {\bibfnamefont {F.~A.}\ \bibnamefont
  {Morrison}} \emph {et~al.},\ }\href@noop {} {\emph {\bibinfo {title}
  {Understanding rheology}}},\ Vol.~\bibinfo {volume} {1}\ (\bibinfo
  {publisher} {Oxford university press New York},\ \bibinfo {year}
  {2001})\BibitemShut {NoStop}%
\bibitem [{\citenamefont {Camp{\`a}s}\ \emph {et~al.}(2014)\citenamefont
  {Camp{\`a}s}, \citenamefont {Mammoto}, \citenamefont {Hasso}, \citenamefont
  {Sperling}, \citenamefont {O'connell}, \citenamefont {Bischof}, \citenamefont
  {Maas}, \citenamefont {Weitz}, \citenamefont {Mahadevan},\ and\ \citenamefont
  {Ingber}}]{campas2014quantifying}%
  \BibitemOpen
  \bibfield  {author} {\bibinfo {author} {\bibfnamefont {O.}~\bibnamefont
  {Camp{\`a}s}}, \bibinfo {author} {\bibfnamefont {T.}~\bibnamefont {Mammoto}},
  \bibinfo {author} {\bibfnamefont {S.}~\bibnamefont {Hasso}}, \bibinfo
  {author} {\bibfnamefont {R.~A.}\ \bibnamefont {Sperling}}, \bibinfo {author}
  {\bibfnamefont {D.}~\bibnamefont {O'connell}}, \bibinfo {author}
  {\bibfnamefont {A.~G.}\ \bibnamefont {Bischof}}, \bibinfo {author}
  {\bibfnamefont {R.}~\bibnamefont {Maas}}, \bibinfo {author} {\bibfnamefont
  {D.~A.}\ \bibnamefont {Weitz}}, \bibinfo {author} {\bibfnamefont
  {L.}~\bibnamefont {Mahadevan}},\ and\ \bibinfo {author} {\bibfnamefont
  {D.~E.}\ \bibnamefont {Ingber}},\ }\bibfield  {title} {\bibinfo {title}
  {Quantifying cell-generated mechanical forces within living embryonic
  tissues},\ }\href@noop {} {\bibfield  {journal} {\bibinfo  {journal} {Nature
  methods}\ }\textbf {\bibinfo {volume} {11}},\ \bibinfo {pages} {183}
  (\bibinfo {year} {2014})}\BibitemShut {NoStop}%
\bibitem [{\citenamefont {Aydin}\ \emph {et~al.}(2019)\citenamefont {Aydin},
  \citenamefont {Rieser}, \citenamefont {Hubicki}, \citenamefont {Savoie},\
  and\ \citenamefont {Goldman}}]{aydin2019physics}%
  \BibitemOpen
  \bibfield  {author} {\bibinfo {author} {\bibfnamefont {Y.~O.}\ \bibnamefont
  {Aydin}}, \bibinfo {author} {\bibfnamefont {J.~M.}\ \bibnamefont {Rieser}},
  \bibinfo {author} {\bibfnamefont {C.~M.}\ \bibnamefont {Hubicki}}, \bibinfo
  {author} {\bibfnamefont {W.}~\bibnamefont {Savoie}},\ and\ \bibinfo {author}
  {\bibfnamefont {D.~I.}\ \bibnamefont {Goldman}},\ }\bibfield  {title}
  {\bibinfo {title} {Physics approaches to natural locomotion: Every robot is
  an experiment},\ }in\ \href@noop {} {\emph {\bibinfo {booktitle} {Robotic
  Systems and Autonomous Platforms}}}\ (\bibinfo  {publisher} {Elsevier},\
  \bibinfo {year} {2019})\ pp.\ \bibinfo {pages} {109--127}\BibitemShut
  {NoStop}%
\bibitem [{\citenamefont {Baek}\ and\ \citenamefont
  {Magda}(2003)}]{baek2003monolithic}%
  \BibitemOpen
  \bibfield  {author} {\bibinfo {author} {\bibfnamefont {S.-G.}\ \bibnamefont
  {Baek}}\ and\ \bibinfo {author} {\bibfnamefont {J.~J.}\ \bibnamefont
  {Magda}},\ }\bibfield  {title} {\bibinfo {title} {Monolithic rheometer plate
  fabricated using silicon micromachining technology and containing miniature
  pressure sensors for n 1 and n 2 measurements},\ }\href@noop {} {\bibfield
  {journal} {\bibinfo  {journal} {Journal of Rheology}\ }\textbf {\bibinfo
  {volume} {47}},\ \bibinfo {pages} {1249} (\bibinfo {year}
  {2003})}\BibitemShut {NoStop}%
\bibitem [{\citenamefont {Khair}\ and\ \citenamefont
  {Squires}(2010)}]{khair2010active}%
  \BibitemOpen
  \bibfield  {author} {\bibinfo {author} {\bibfnamefont {A.~S.}\ \bibnamefont
  {Khair}}\ and\ \bibinfo {author} {\bibfnamefont {T.~M.}\ \bibnamefont
  {Squires}},\ }\bibfield  {title} {\bibinfo {title} {Active microrheology: a
  proposed technique to measure normal stress coefficients of complex fluids},\
  }\href@noop {} {\bibfield  {journal} {\bibinfo  {journal} {Physical review
  letters}\ }\textbf {\bibinfo {volume} {105}},\ \bibinfo {pages} {156001}
  (\bibinfo {year} {2010})}\BibitemShut {NoStop}%
\bibitem [{\citenamefont {Elfring}\ and\ \citenamefont
  {Lauga}(2015)}]{elfring2015theory}%
  \BibitemOpen
  \bibfield  {author} {\bibinfo {author} {\bibfnamefont {G.~J.}\ \bibnamefont
  {Elfring}}\ and\ \bibinfo {author} {\bibfnamefont {E.}~\bibnamefont
  {Lauga}},\ }\bibfield  {title} {\bibinfo {title} {Theory of locomotion
  through complex fluids},\ }in\ \href@noop {} {\emph {\bibinfo {booktitle}
  {Complex fluids in biological systems}}}\ (\bibinfo  {publisher} {Springer},\
  \bibinfo {year} {2015})\ pp.\ \bibinfo {pages} {283--317}\BibitemShut
  {NoStop}%
\bibitem [{\citenamefont {Walters}\ and\ \citenamefont
  {Waters}(1963)}]{walters1963use}%
  \BibitemOpen
  \bibfield  {author} {\bibinfo {author} {\bibfnamefont {K.}~\bibnamefont
  {Walters}}\ and\ \bibinfo {author} {\bibfnamefont {N.}~\bibnamefont
  {Waters}},\ }\bibfield  {title} {\bibinfo {title} {On the use of a rotating
  sphere in the measurement of elastico-viscous parameters},\ }\href@noop {}
  {\bibfield  {journal} {\bibinfo  {journal} {British Journal of Applied
  Physics}\ }\textbf {\bibinfo {volume} {14}},\ \bibinfo {pages} {667}
  (\bibinfo {year} {1963})}\BibitemShut {NoStop}%
\bibitem [{\citenamefont {Walters}\ and\ \citenamefont
  {Waters}(1964{\natexlab{a}})}]{walters1964interpretation}%
  \BibitemOpen
  \bibfield  {author} {\bibinfo {author} {\bibfnamefont {K.}~\bibnamefont
  {Walters}}\ and\ \bibinfo {author} {\bibfnamefont {N.}~\bibnamefont
  {Waters}},\ }\bibfield  {title} {\bibinfo {title} {The interpretation of
  experimental results obtained from a rotating-sphere elastoviscometer},\
  }\href@noop {} {\bibfield  {journal} {\bibinfo  {journal} {British Journal of
  Applied Physics}\ }\textbf {\bibinfo {volume} {15}},\ \bibinfo {pages} {989}
  (\bibinfo {year} {1964}{\natexlab{a}})}\BibitemShut {NoStop}%
\bibitem [{\citenamefont {Walters}\ and\ \citenamefont
  {Waters}(1964{\natexlab{b}})}]{walters1964steady}%
  \BibitemOpen
  \bibfield  {author} {\bibinfo {author} {\bibfnamefont {K.}~\bibnamefont
  {Walters}}\ and\ \bibinfo {author} {\bibfnamefont {N.}~\bibnamefont
  {Waters}},\ }\bibfield  {title} {\bibinfo {title} {The steady flow of a
  rivlin-ericksen fluid induced by a rotating-sphere},\ }\href@noop {}
  {\bibfield  {journal} {\bibinfo  {journal} {Rheologica Acta}\ }\textbf
  {\bibinfo {volume} {3}},\ \bibinfo {pages} {312} (\bibinfo {year}
  {1964}{\natexlab{b}})}\BibitemShut {NoStop}%
\bibitem [{\citenamefont {Fattal}\ and\ \citenamefont
  {Kupferman}(2004)}]{Fattal2004}%
  \BibitemOpen
  \bibfield  {author} {\bibinfo {author} {\bibfnamefont {R.}~\bibnamefont
  {Fattal}}\ and\ \bibinfo {author} {\bibfnamefont {R.}~\bibnamefont
  {Kupferman}},\ }\bibfield  {title} {\bibinfo {title} {{Constitutive laws for
  the matrix-logarithm of the conformation tensor}},\ }\href
  {https://doi.org/10.1016/j.jnnfm.2004.08.008} {\bibfield  {journal} {\bibinfo
   {journal} {Journal of Non-Newtonian Fluid Mechanics}\ }\textbf {\bibinfo
  {volume} {123}},\ \bibinfo {pages} {281} (\bibinfo {year}
  {2004})}\BibitemShut {NoStop}%
\bibitem [{\citenamefont {Hulsen}\ \emph {et~al.}(2005)\citenamefont {Hulsen},
  \citenamefont {Fattal},\ and\ \citenamefont {Kupferman}}]{Hulsen2005}%
  \BibitemOpen
  \bibfield  {author} {\bibinfo {author} {\bibfnamefont {M.~A.}\ \bibnamefont
  {Hulsen}}, \bibinfo {author} {\bibfnamefont {R.}~\bibnamefont {Fattal}},\
  and\ \bibinfo {author} {\bibfnamefont {R.}~\bibnamefont {Kupferman}},\
  }\bibfield  {title} {\bibinfo {title} {{Flow of viscoelastic fluids past a
  cylinder at high Weissenberg number: Stabilized simulations using matrix
  logarithms}},\ }\href {https://doi.org/10.1016/j.jnnfm.2005.01.002}
  {\bibfield  {journal} {\bibinfo  {journal} {Journal of Non-Newtonian Fluid
  Mechanics}\ }\textbf {\bibinfo {volume} {127}},\ \bibinfo {pages} {27}
  (\bibinfo {year} {2005})}\BibitemShut {NoStop}%
\bibitem [{\citenamefont {Shaqfeh}\ \emph {et~al.}()\citenamefont {Shaqfeh},
  \citenamefont {Prakash}, \citenamefont {Kroo},\ and\ \citenamefont
  {Binagia}}]{shaqfeh_prakash_kroo_binagia}%
  \BibitemOpen
  \bibfield  {author} {\bibinfo {author} {\bibfnamefont {E.~S.}\ \bibnamefont
  {Shaqfeh}}, \bibinfo {author} {\bibfnamefont {M.}~\bibnamefont {Prakash}},
  \bibinfo {author} {\bibfnamefont {L.~A.}\ \bibnamefont {Kroo}},\ and\
  \bibinfo {author} {\bibfnamefont {J.}~\bibnamefont {Binagia}},\ }\href@noop
  {} {\bibinfo {title} {Patent application: A mechanical swimmer that acts as a
  rheometer}},\ \bibinfo {note} {application Number: 63/230448, Filed Aug 06,
  2021, Assignee: The Board of Trustees of the Leland Stanford Junior
  University}\BibitemShut {NoStop}%
\end{thebibliography}%

\newpage

\section*{Supplemental Materials}

\subsection*{Swimmer Motion Tracking}

To obtain rotation speed of the small sphere, a motion tracking algorithm was applied to videos taken along the axis of rotation. Equally spaced meridional lines were painted on the sphere to facilitate tracking. A circular Hough transform (CHT) was used to find the sphere in the image, then position data for each line was extracted, and an optimization problem solved to find the new rotational position of the sphere from frame to frame. The mechanical control of rotation was verified to be constant in rotational speed and consistent from trial to trial (Fig A.1?) 

For the extraction of translational motion, a similar object tracking scheme was used with video data taken orthogonal to the axis of movement. The spheres were found using a CHT in each frame of interest, then the center of mass calculated from known masses of each sphere and the linker. The distance travelled in each frame was calculated from pixel distance by applying a transform using the known radius of each sphere. Due to the very small forward speed, the average velocity for each trial was calculated with a linear best-fit regression. The domain of each regression was chosen to be between the startup regime and before any boundary effects influenced the motion. 

Exceptional linear regression fits ($R^2 >0.99$) demonstrate the reliability of the stepper-motor based speed control system in its ability to maintain consistent rotational speed control over many minutes, and over a wide range of speed settings. 

\begin{figure}[ht!]
    \centering
    \includegraphics[width = 0.45\textwidth]{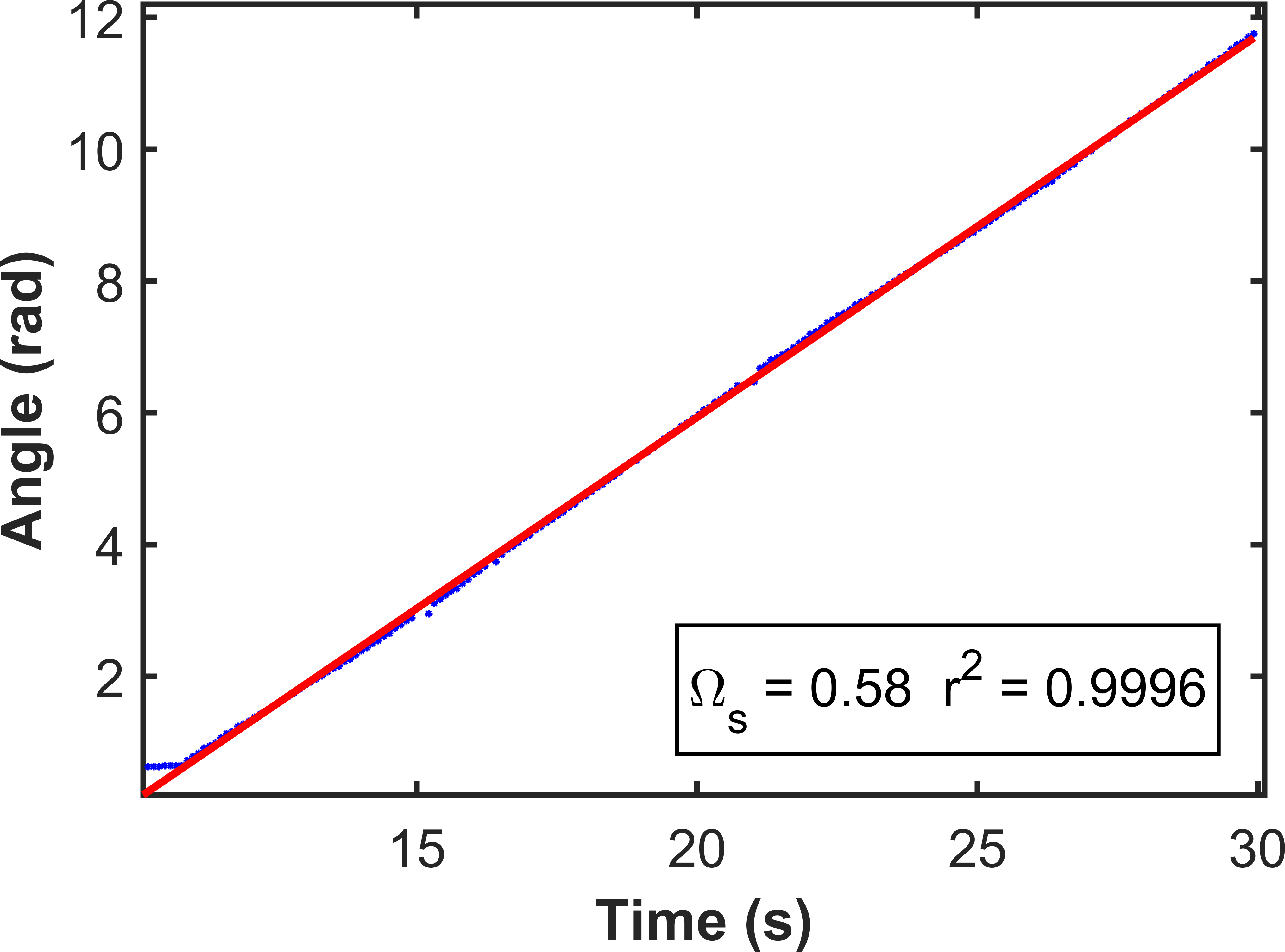}
    \caption{Rotation tracking example}
    \label{fig:my_label}
\end{figure}

\begin{figure}[ht!]
    \centering
    \includegraphics[width = 0.45\textwidth]{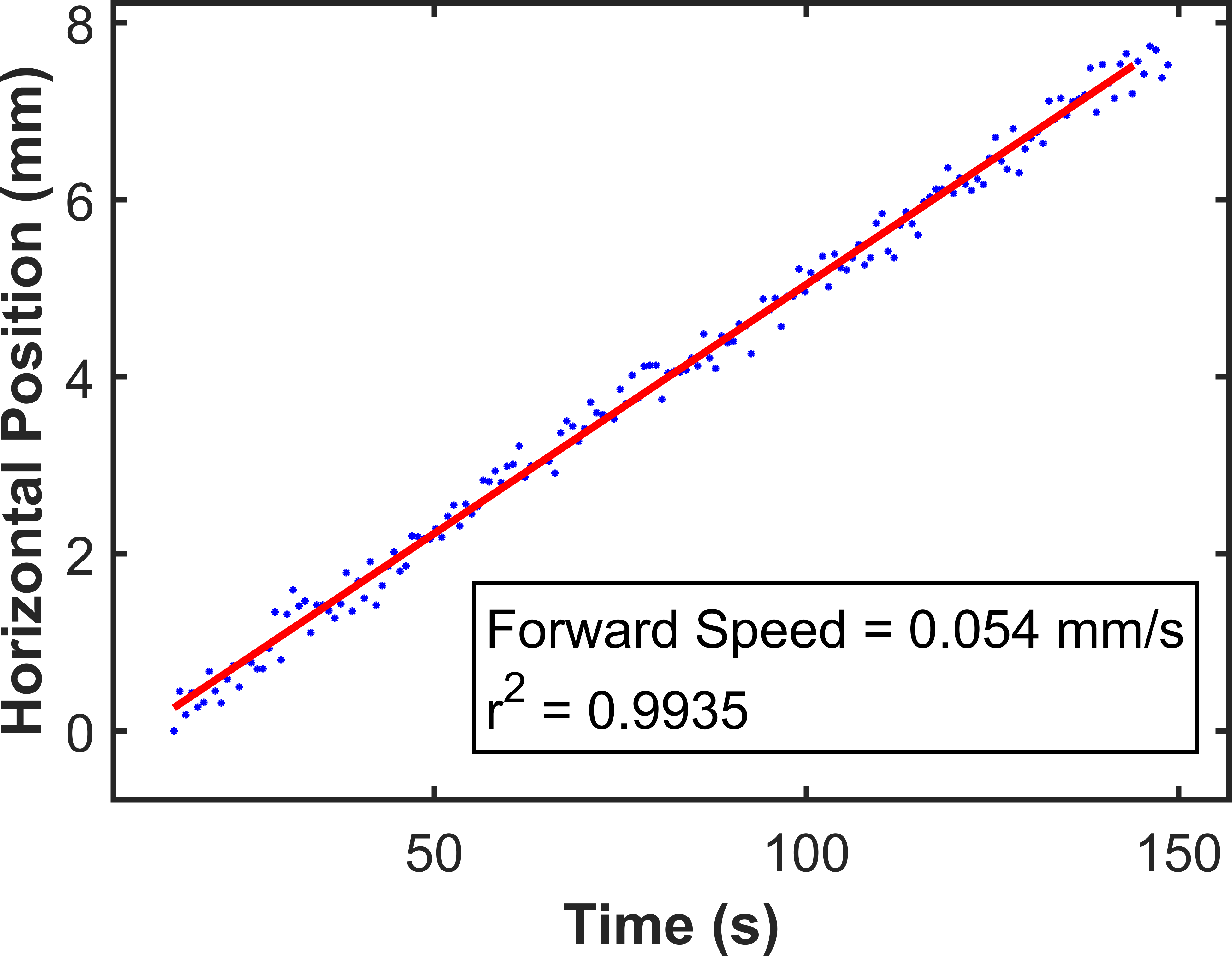}
    \caption{Translation tracking example}
    \label{fig:my_label}
\end{figure}

\end{document}